\documentclass[useAMS,usenatbib]{mn2e}

\usepackage{graphicx}
\usepackage{natbib}
\usepackage{amsmath}
\def\simlt{\mathrel{\rlap{\lower 3pt\hbox{$\sim$}}\raise 2.0pt\hbox{$<$}}}
\def\simgt{\mathrel{\rlap{\lower 3pt\hbox{$\sim$}} \raise 2.0pt\hbox{$>$}}}
\def\lsim{\mathrel{\rlap{\lower 3pt\hbox{$\sim$}}\raise 2.0pt\hbox{$<$}}}
\def\gsim{\mathrel{\rlap{\lower 3pt\hbox{$\sim$}} \raise 2.0pt\hbox{$>$}}}

\def\di{\mbox{d}}
\def\Msun{{\rm M}_{\odot}}
\def\Zsun{{\rm Z}_{\odot}}
\def\Mpch{Mpc/{\it h}}
\def\Msunh{M$_\odot$/{\it h}}

\begin{document}

\title[Simulating high-$z$ GRB host galaxies]{Simulating high-$z$ Gamma-ray Burst host galaxies}

\author[Salvaterra et al.]{R.~Salvaterra$^1$, U.~Maio$^2$, B.~Ciardi$^3$, M.~A.~Campisi\\
$^1$ INAF, IASF Milano, via E. Bassini 15, I-20133 Milano, Italy \\
$^2$ Max-Planck-Institut f\"ur extraterrestrische Physik,
Giessenbachstra\ss e 1, D-85748 Garching bei M\"unchen, Germany\\
$^3$ Max-Planck-Institut f\"ur Astrophysik, Karl-Schwarzschild-Stra\ss e 1, D-85748 Garching bei M\"unchen, Germany\\
}

\maketitle \vspace {7cm}

\begin{abstract}
We investigate the nature of high-$z$ host galaxies of long Gamma-Ray Bursts (LGRBs) by means of state-of-the-art numerical simulations of cosmic structure formation and evolution of galaxies.
We combine results from different runs with various box sizes and
resolutions. 
By assigning to each simulated galaxy the probability to host
a LGRB, assumed to be proportional to the mass of young stars,
we provide a full description of the physical properties of high-$z$
LGRB host galaxy population.
We find that LGRBs at $z>6$ are hosted in galaxies with typical star
formation rates $\rm SFR\simeq 0.03-0.3\;\Msun$ yr$^{-1}$, stellar
masses $M_\star\simeq 10^6-10^8~\Msun$, and metallicities $Z\simeq
0.01-0.1\;\Zsun$. 
Furthermore, the ratio between their doubling time and the
corresponding cosmic time seems to be universally equal to $\sim 0.1-0.3$, independently from the redshift.
  The distribution of their UV luminosity places LGRB hosts in the faint-end of the galaxy luminosity function, well below the current capabilities of space- or ground-based optical facilities. This is in line with recent reports of non-detection of LGRB hosts using extremely deep HST and VLT observations. In conclusion, high-$z$ LGRBs are found to trace the position of those faint galaxies that are thought to be the major actors in the re-ionization of the Universe.
\end{abstract}


\section{Introduction} \label{sect:introduction}

Long Gamma-Ray Bursts (LGRBs) are powerful flashes of $\gamma$-rays that are observed with a frequency of about one per day over the whole sky.
The $\gamma$-ray emission is accompanied by a long-lasting tail, called afterglow, usually detected over the whole electromagnetic spectrum.
Their extreme brightness easily over-shines the luminosity of their host galaxy and makes them detectable up to extremely high redshifts, as shown by the discovery of GRB~090423 at $z=8.2$ \citep{sal09,tan09}, or of GRB~090429B at $z \sim 9.4$ \citep{Cucchiara2011}.
Metal absorption lines can often be identified in their afterglow spectra, allowing a study of the properties of the environment in which they blow.
Once the afterglow has faded, follow-up searches of the LGRB host galaxy become possible.

Observations of high-$z$ LGRBs can provide unique information about the high-redshift Universe \cite[see][and references therein]{McQuinn2009}.
For example, LGRBs can be used to measure the neutral-hydrogen fraction
of the intergalactic medium \cite[e.g.][]{McQuinn2008, Gallerani2008,
  Nagamine2008}, to study the galaxy metal and dust content
\cite[e.g.][]{Savaglio2012,Wang2012}, to probe the intergalactic
radiation field \cite[e.g.][]{Inoue2010} or even to constrain the level of primordial
non-Gaussianities \citep{Maio2012}.
The tight observational link between high-$z$ distant galaxies and LGRBs is particularly evident when looking at their conditions during the first few  billion years of life of the Universe. In fact, for the first time, surprisingly high metallicities (with super-solar abundances) have been estimated from GRB~090323 at redshift $z \sim 3.5$, in correspondence of a merging event between two galaxies \citep{Savaglio2012}.
The significant star formation rate ($\sim 6\,\rm \Msun/yr$) enhanced by their interaction might probably have triggered the birth of the LGRB progenitor and the simultaneous determination of a disturbed, metal-rich system, identifiable as a high-redshift, massive, sub-millimeter galaxy.
\\
Moreover, LGRBs might represent the most promising way to directly detect the very first stars: the so-called Population III (Pop III) stars \cite[e.g.][]{ciardi00,toma10,campisi2011,deSouza2011}.
These form in pristine environments and are predicted to have large masses ($\sim 10-10^2\,\rm M_\odot$) \cite[e.g.][]{woosley2002,heger2003,woo06b}, because the most relevant coolants in such environments (H-derived molecules, like H$_2$ or HD) are not able to cool the gas efficiently and induce small-scale fragmentation.
This is, instead, possible only in metal-enriched gas, where common Population II-I (Pop II-I) stars can form  (\citealt{baraffe2001}), either via low-temperature fine-structure transition metal cooling 
\cite[e.g.][when a minimum critical metallicity $Z_{crit}\sim
10^{-4}\Zsun$ is reached]{Schneider2002,Bromm2003}, or via dust
cooling (with a minimum critical metallicity as low as $Z_{crit}\sim 10^{-6}\Zsun$, according to \citealt{Schneider2003,Schneider2012}).
The different initial mass functions (IMF) for Pop~III stars can influence the thermodynamic state of the cosmic medium and the evolution of the following Pop~II regime.
Indeed, larger-mass stellar populations have harder spectral energy distribution (SED) and can instantaneously produce more photons \citep{Schaerer2002}, efficiently heating the cosmic gas.
Also, they die and enrich the surrounding medium earlier, leading more rapidly (in any case, within $\sim 10^7-10^8\,\rm yr$) to the following Pop~II star formation regime \cite[e.g.][]{maio2010}.
Furthermore, since different progenitor stars (i.e., massive Pop II or Pop III stars) can possibly affect the resulting LGRB rate \cite[][]{campisi2011}, to investigate the LGRBs properties it is crucial to have a good description of early structure formation, and 
to properly model the transition from the primordial, Pop III regime to the 
following, standard, Pop II one 
\citep{tornatore2007, maio2010, maio2011, wise2012}.

In this paper, we will use state-of-the-art N-body, hydrodynamic,
chemistry simulations \cite[][]{maio2010} of structure formation and
evolution of high-$z$ galaxies to unveil the nature and the physical
properties of Pop II LGRB hosts at early epochs. 
Our work extends toward  higher redshifts the
 results already obtained by 
previous theoretical studies of the nature of LGRB hosts
\citep{Courty2004,Nuza2007,Campisi2009,Chisari2010,campisi2011b} that have
mainly focused on the expected properties of these galaxies at low or
intermediate redshift where models can be compare directly
with observations. In spite of the different assumptions and
approaches,  all these works predict that LGRB should be
hosted in low-mass, young, star-forming galaxies, consistently with observations \cite[see
e.g.][]{Savaglio2009}. As this galaxy population is believed to
dominate the star formation activity in the early Universe
\citep{Salvaterra2010, Jaacks2012,Munoz2011,Lacey2011}, we would like to investigate to which extend LGRB can be
used as a new tool to study the first phases of structure formation. 

Furthermore,   
extensive searches of high-$z$ LGRB host galaxies has been recently carried out with HST \citep{Tanvir2012} and VLT \citep{Basa2012}.
In spite of the very deep limits reached by these observations, none of the targets have been identified.
The lower limits on the UV luminosities reached using the deepest
observations available correspond to magnitudes $\rm M_{UV} \sim
-17\div -20 $, 
as precisely reported in Table~\ref{tab:obs} together with upper
limits for the star formation rate (SFR).  These limits suggest
  that high-$z$ LGRB hosts have lower SFRs with respect to
  their low-$z$ counterparts \citep{Basa2012}. By comparing available
  observation limits to the results of our numerical simulations,
  we will study the detectability of high-$z$ LGRB hosts with current
  and future instruments. 
We will include in our analysis
GRB~060927 at $z=5.47$ \citep{Ruiz07},
and the four known LGRBs at $z>6$:
GRB~050904 at $z=6.3$ \citep{kaw06},
GRB~080913 at $z=6.7$ \citep{Greiner09},
GRB~090423 at $z=8.2$ \citep{sal09,tan09}, and
GRB~090429B with a photometric redshift of $z\sim 9.4$ 
\citep{Cucchiara2011} -- see also Table~\ref{tab:obs}.
We will also provide the expected LGRB host distributions as function
of stellar masses ($M_\star$), gas phase metallicities ($Z$), and specific SFR (sSFR).
This will give us a complete theoretical picture of the typical
environments where Pop II LGRBs should be found at redshift $z\sim 6 -
10$, as inferred from the analyses of our numerical simulations.
Regarding LGRBs powered by massive Pop III stars, we refer the
interested reader to our previous work, \cite{campisi2011}, where we
compute their rate and the properties of their host galaxies by means
of the same numerical simulations used here.

The paper is organized as follows:
in Section~\ref{sect:simulations} we present the N-body, hydrodynamical, chemistry simulations used in this work;
in Section~\ref{sect:GRBhosts} we compare the simulated host properties with available data;
and, finally, we summarize our results in Section~\ref{sect:conclusions}.
Throughout the paper magnitudes are given in the AB system, logarithms
are always base-10, and the underlying cosmological model assumed for
all the calculations and simulations is a standard $\Lambda$CDM model,
with total-matter density parameter at the present $\Omega_{0,m}=0.3$,
cosmological constant density parameter $\Omega_{0,\Lambda}=0.7$,
baryonic-matter density parameter $\Omega_{0,b}=0.04$, expansion rate
in units of 100~km/s/Mpc $h = 0.7$, spectral normalization
$\sigma_8=0.9$, and primordial spectral index $n=1$. Unless
  differently specified, metallicities refer to the gas phase.

\begin{table}
\centering
\begin{tabular}{lccc}
\hline
GRB & $z$ & M$_{UV}$ & SFR [$\Msun$/yr] \\
\hline
060927 & 5.47 & $>-18.02$ & $<0.65$ \\
050904 & 6.3 & $>-19.95$ & $<4.1$ \\
080913 & 6.7 & $>-19.00$ & $<1.3$ \\
090423 & 8.2 & $>-16.95$ & $<0.38$ \\
090429B & $\sim 9.4$ & $>-19.65$ & $<2.4$ \\
\hline
\end{tabular}
\caption{
High-$z$ LGRB host observations from Tanvir et al. (2012) which are
considered in this work. The columns refer to, from left to right:
name of the LGRB, its redshift $z$, UV magnitude M$_{UV}$ and
star formation rate.
}
\label{tab:obs}
\end{table}

\section{Simulations and analysis} \label{sect:simulations}

In the next sections, we will briefly describe the essential features of our simulations (Section~\ref{sect:sims}), and then we will present the data analysis (Section~\ref{sect:an}) leading to the results shown in the following Section~\ref{sect:GRBhosts}.

\subsection{Simulations}\label{sect:sims}

\begin{table*}
\centering
\begin{tabular}{cccccc}
\hline
Model & $L$ [\Mpch] & $m_{\rm gas}$ [\Msunh]& $m_{\rm DM}$ [\Msunh] & $\eta$ [kpc/{\it h}]& log(SFR[$\rm\Msun yr^{-1}$])\\
\hline
L30 & 30  & $9\times 10^6$ & $6\times 10^7$ & 4.7 & $>0$ \\   
L10 & 10  & $3\times 10^5$       & $2\times 10^6$ & 1.0 & $[-1.5,0]$\\   
L5  & 5   & $4\times 10^4$       & $3\times 10^5$ & 0.5 & $<-1.5$ \\   
\hline
\end{tabular}
\caption{Simulation parameters. The columns refer to, from left to right: model name; comoving box length, $L$;
  gas particle mass, $m_{\rm gas}$; dark matter particle mass, $m_{\rm DM}$;
  softening length, $\eta$; star formation rate, $\dot{M_*}$.}
\label{tab:sims}
\end{table*}

\noindent
Here we use the numerical simulations by \cite{maio2010}, in which gas collapse and condensation through atomic or molecular cooling \cite[e.g.][]{GalliPalla1998,abel2002,yoshida2003,maio2006,maio2007,maio2009,mkc2011, wise2007,wise2012} is followed down to proto-galactic scales, where star formation activity \cite[e.g.][]{katz1992,hernquist1989,hernquist2003,springel2003} takes place and induces feedback effects on the ambient medium.
Supernova (SN) explosions distribute metals in the surrounding gas \citep[as described in e.g.][]{springel2003, tornatore2003, tornatoreBDM2007, tornatore2010} altering its chemical composition on the short timescales typical of massive stars \cite[][]{woosley2002,heger2002,Schaerer2002}.

More in detail, for each stellar particle we follow its timescales
\citep{Padovani1993} and yields
\citep{woosley1995,heger2002,heger2010,thielemann2003,vandenhoek1997}
according to a given IMF (see later).
When SNe take place the surrounding gas is randomly 'kicked' with a velocity of $\sim 500\,\rm km/s$ (kinetic feedback) and heated up at temperatures of roughly $\sim 10^5\rm K$ (thermal feedback).
Simultaneously, different metal species (C, O, Si, Fe, Mg) are spread around the neighbouring regions according to the corresponding yields, by mean of a kernel-based smoothing procedure that mimics metal diffusion\footnote{
We note that metal diffusion is a complex issue and currently cannot be properly treated neither in SPH codes (because of the difficulty in diffusing the material above the smoothing scale), nor in grid codes (that are affected by over-mixing problems).
}.
Dependencies on numerical resolution of the different sub-grid models are minor, as also studied in, e.g., \cite{maioiannuzzi2011, maio2011}, and they usually affect only the very first bursts of star formation, to converge relatively soon at slightly later times.
Further effects of different IMFs, SN rages, yields determinations, or possible changes in the basic SPH implementation have been discussed extensively in \cite{maio2010}.

The transition between the Pop III and Pop II star formation regime mentioned in the Introduction is assumed to happen at a critical metallicity $Z_{crit}=10^{-4}\Zsun$, such that if the star forming gas has $Z<Z_{crit}$, then a Salpeter-like Pop III IMF is assumed with mass range between $100\,\rm M_\odot$ and $500\,\rm M_\odot$ and a pair-instability SN range between $140\,\rm M_\odot$ and $260\,\rm M_\odot$ \citep{woosley2002}; otherwise, a standard Salpeter IMF is adopted with mass range between $0.1\,\rm M_\odot$ and $100\,\rm M_\odot$, and SNII range between $8\,\rm M_\odot$ and $40\,\rm M_\odot$.
\\
The cosmological field is sampled for dark matter and baryonic matter at redshift $z=100$, following a common Gaussian matter distribution\footnote{
Deviations from Gaussianity are not expected to significantly alter our conclusions \cite[see][]{maioiannuzzi2011,maiocqg2011,maiokhochfar2012}.
}
.
In particular, here we consider three cubic volumes with comoving sides of 30~\Mpch, 10~\Mpch, and 5~\Mpch.
The identification of the simulated objects (with their gaseous, dark
and stellar components) was carried out by applying a
friends-of-friends technique, with comoving linking length of $20\%$
the mean inter-particle separation, and a minimum number of 32
particles. To ensure and resolve properly the low-mass end of our
  simulated objects, we will consider only well-resolved structures
  with at least 100 baryon particles\footnote{\cite{Bate1997} have shown that objects containing 3-4 times the number of neighbors on which density is computed are properly resolved. In our simulations, 32 neighbors are considered so that the cut at $> 100$ baryon particles is a reasonable choice to avoid numerical problems.} (the minimum number of dark matter
  particles per object is usually an order of magnitude larger). Moreover, we check that our results do not change
  significantly 
  if only objects with more than 300 baryon particles are considered. Substructures are identified by using a SubFind algorithm, which follows a spherical-overdensity approach and additionally discriminates among bound and non-bound particles \cite[e.g.][]{Dolag2009}.
The parameters used in the simulations are listed in Table~\ref{tab:sims}.
For more details we refer the reader to the original paper \citep{maio2010}.
\\
Before discussing the analysis of the simulations, it should be
noticed that, despite the amount of detailed physical implementations included, some ingredients are still missing or could be further improved.
For example, metal diffusion is accounted for by smoothing over the SPH kernel: this approach might work reasonably on small scales, but it will probably have some limitations on large scales.
Neither radiative feedback \citep[e.g][]{gnedinabel2001,whalen2008}
nor thermal conduction \citep[e.g.][]{dolag2004}
are included, and both are likely to introduce some modifications in the properties of the host galaxies.
On the other hand, other specific issues, like numerical-viscosity
schemes or smoothing kernels to refine the SPH algorithms
are likely to give major improvements when resolving the details of
hydrodynamical shocks and turbulence, but they should not affect
significantly the general trends we find here \citep[][and references therein]{price2012b}.

\subsection{Simulation analysis}\label{sect:an}

\begin{figure}
\center{\includegraphics[scale=0.44]{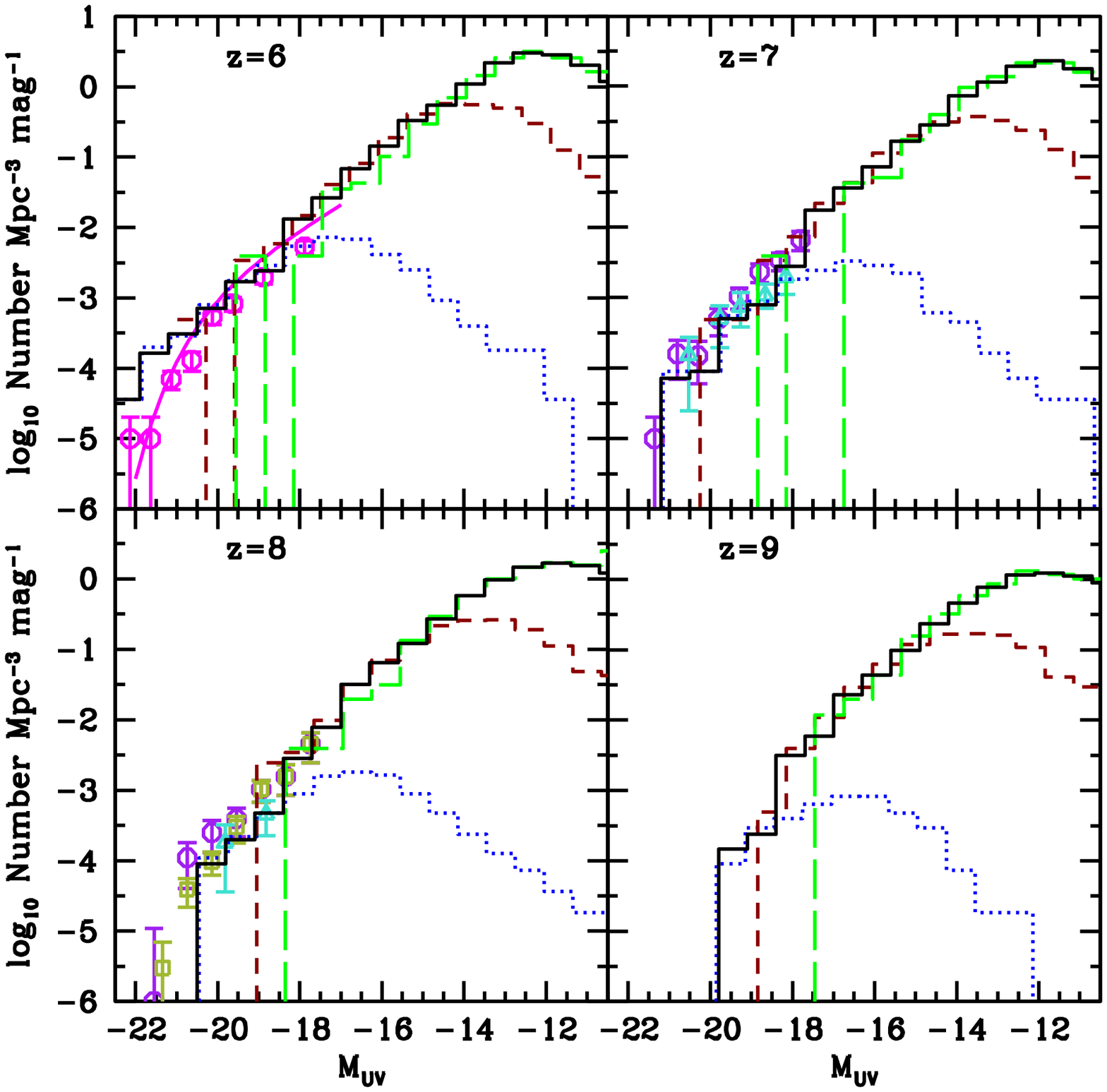}}
\caption{\label{fig:LF} Luminosity functions at $z=6$, 7, 8 and 9. 
 Long-dashed (green), short-dashed (red) and dotted (blue) histograms
 show the result of L5, L10 and L30, respectively, while solid (black)
 histograms refer to the luminosity functions obtained using jointly
 the three simulations. Data are from: \citet[][circles]{Bouwens2007} 
 at $z=6$; \citet[][circles]{Bouwens2011}  and \citet[][triangles]{McLure}
 at $z=7$, 8; \citet[][squares]{Oesch2012}  at $z=8$.  The
continuos line at $z=6$ is the fit to the LF as obtained by \citet[][]{Su2011}.
}
\end{figure}

Rare galaxies with very high SFRs are properly accounted for only in the larger simulation box, while galaxies at very low SFR are resolved only in the small simulation box which has a better numerical resolution.
Therefore, we jointly consider the results of different simulation boxes by selecting galaxies on the basis of their SFRs, which are reported in the last column of  Table~\ref{tab:sims}.
The SFR thresholds are identified by inspecting the number density
distributions and correspond to the SFR bins in which the galaxy
number density is consistent in the larger and smaller box. We check at which level both numerical resolution and the unavoidable lack of
smaller, unresolved, star forming structures in the larger boxes \citep{nag04} alter our results going from the smaller to the larger simulation box.
We find that, in the mass-SFR plane, SFRs roughly agree when considering objects with the same stellar masses extracted from the three different boxes.
However, we obtain a clear, constant
displacement in the gas phase metallicities of galaxies with the same mass
(or SFR) going from L5 to L10 and L30 suggestive of a numerical resolution effect \citep{nag04}.
Therefore, we renormalize the metallicities of simulated objects by adding 0.1 dex and 0.5 dex to all galaxies (at all redshifts) in L10 and L30, respectively.
We notice that since most of the LGRBs ($\sim 90$\%) form in objects selected from L5 and L10, the rather large correction applied to metallicities in L30 will not affect significantly our final results (see later).

The total luminosity of a galaxy at wavelength $\lambda$, $L_\lambda$
(in  $\rm erg~s^{-1}~Hz^{-1}$), forming Pop~II stars at a rate SFR$^{\rm II}$ and Pop~III stars at a rate SFR$^{\rm III}$, is then given by\footnote{We notice that, despite their similar notations, in the following $l_{\lambda}^{\rm II}$ and $l_{\lambda}^{\rm III}$ are dimensionally two different quantities.}

\begin{equation}
L_\lambda 
= L_{\lambda}^{\rm II} + L_{\lambda}^{\rm III} 
= l_{\lambda}^{\rm II}\left(\tau^{\rm II},Z\right){\rm SFR}^{\rm II} 
+ l_{\lambda}^{\rm III}{\rm SFR}^{\rm III}\tau^{\rm III},
\end{equation}
where $l_{\lambda}^{\rm II}\left(\tau^{\rm II},Z\right)$ is the SED template (in $\rm erg~s^{-1}~Hz^{-1}~M_\odot^{-1}~yr $) corresponding to the energy emitted per unit time and frequency ($\rm erg~s^{-1}~Hz^{-1}$), and per unit star formation rate (in $\rm \Msun~ yr^{-1}$) for Pop~II stars with mean age $\tau^{\rm II}$ and metallicity $Z$.
Similarly, $l_{\lambda}^{\rm III}$ is the SED template (in $\rm
erg~s^{-1}~Hz^{-1}~M_\odot^{-1} $) corresponding to the energy emitted
per unit time and frequency ($\rm erg~s^{-1}~Hz^{-1}$), and per unit
stellar mass (in $\rm \Msun$) for Pop~III stars with mean lifetime
$\tau^{\rm III} = 2.5\times 10^6~\rm yr$ \citep{Schaerer2002}. We implicitly assume that the emission properties of Pop III stars are roughly constant during their short lifetime.
Finally, we neglect dust absorption in our calculation. As shown by \cite{Salvaterra2010}, dust extinction for galaxies with $M_\star<10^8\;\Msun$ at $z>6$ is $E(B-V)<0.01$, while a larger contribution is expected in more massive and metal-rich objects \cite[e.g.][]{Dayal2010}. As it will be clear in the next Section, the large majority of LGRBs are hosted in low-mass galaxies so that our assumption of no dust will not affect our conclusions.

As a consistency check, in Fig.~\ref{fig:LF} we show the luminosity function (LF) of galaxies at $z=6$, 7, 8 and 9. The solid histogram reports the LF as obtained 
by considering jointly the three simulation boxes. The resulting LF is
compared with data from the literature.  In the range
  of luminosities in which observational data and simulation results
  overlap, we find a good agreement with the amplitude of the observed
  galaxy  LF at all redshifts.  Moreover, 
  our simulations, similarly to other recent numerical
  \citep{Salvaterra2010,Jaacks2012} and semi-analytical
  \citep{Munoz2011,Lacey2011} studies, predict a rather steep
  faint-end slope consistent with the estimates based on HUDF data \citep{Bouwens2011}. The agreement is
  even more remarkable as no attempts have been made to fit or adjust
  the theoretical curves to the observed LF, i.e. they have been
  computed directly from the simulation output with no free extra
  parameters.  We note that the bright-end of the LF 
at $z=6$ is slightly overestimated by our simulations, in particular by L30. This effect is  
indeed expected since, as mentioned above, we neglect dust extinction important for bright massive objects.
We note however that this does not affect the conclusions of our paper, as less than 5\% of LGRBs are 
found to be hosted in galaxies brighter than $M_{UV}=-20$ (see
Fig.~\ref{fig:Muv}).
In conclusion, our simulations provide a good description of the galaxy
population in the early Universe in the luminosity range probed by HST and match the
extrapolation of the observed faint-end slope in the redshift range we
are interested in. In the following, we will make use of these results
to predict the
nature and the properties of LGRB host galaxies in the redshift range
$z=6-10$.

\begin{figure*}
  \centering
  \hspace{-1.2cm}
  \includegraphics[scale=0.46]{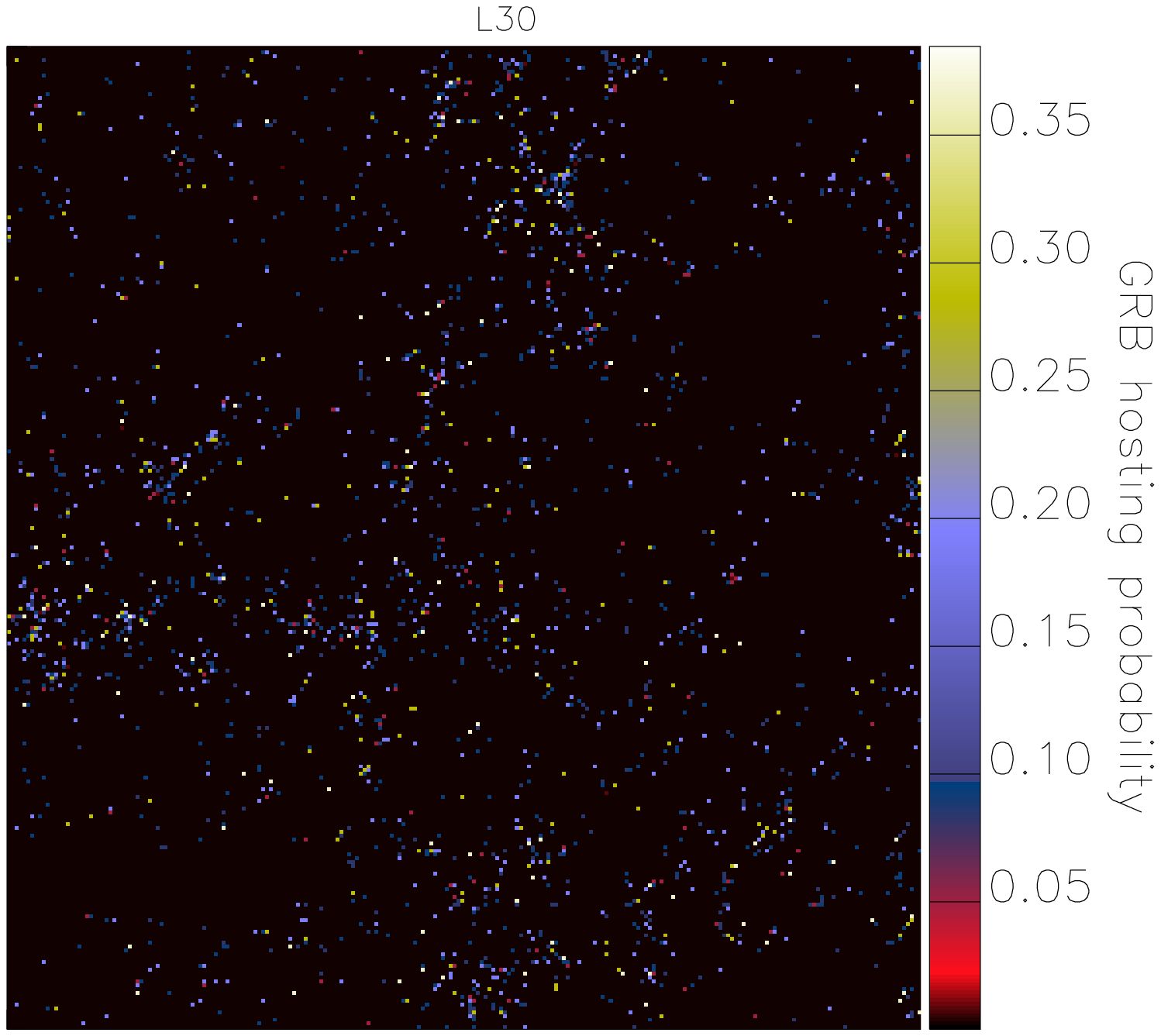} \hspace{-1.97cm}
  \includegraphics[scale=0.46]{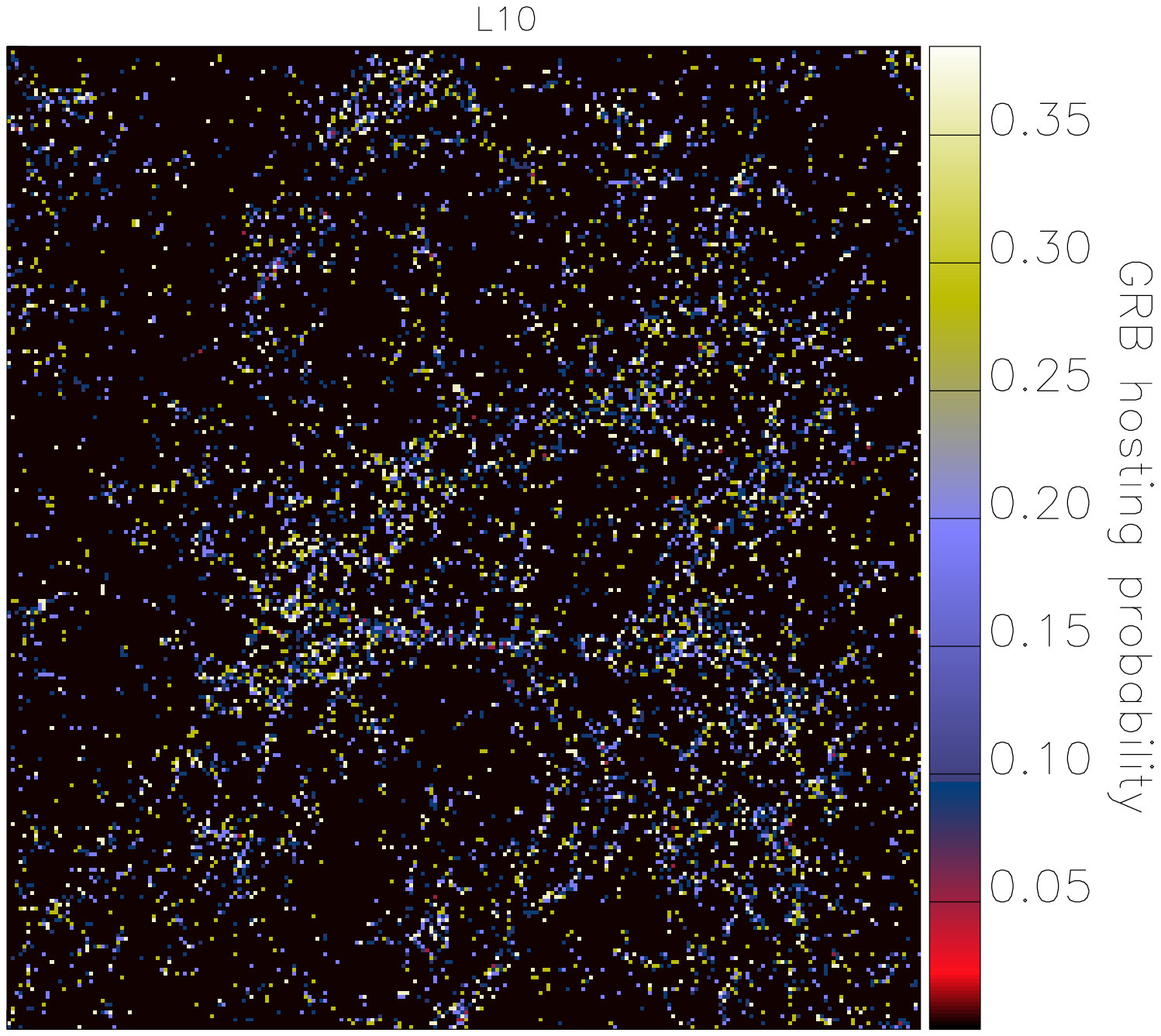} \hspace{-1.97cm}
  \includegraphics[scale=0.46]{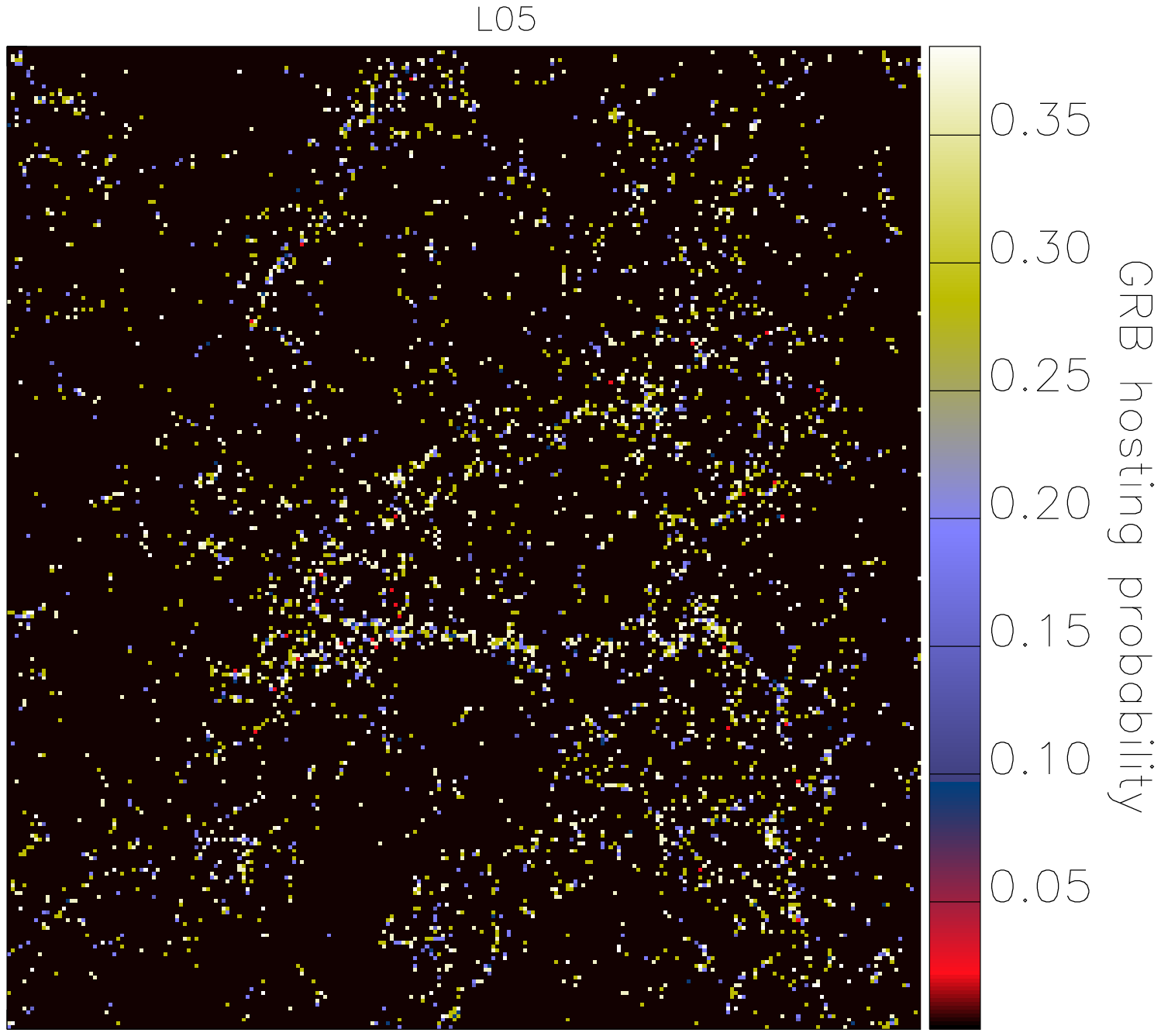}\\
  \caption{
    Differential probability of finding LGRBs per logarithmic SFR bin in L30 (left panel), L10 (central) and L5 (right). Each color pixel refers to a galaxy and is color coded according to the probability in eq.~(\ref{eq:dPdLogSFR}).
  }
  \label{fig:mappedPGRB}
\end{figure*}

\section{Simulated LGRB host galaxies}\label{sect:GRBhosts}

Long GRBs are believed to be linked to the death of massive stars
\cite[e.g.][]{Fryer_Woosley_Hartmann_1999,woo06b}. 
This idea is supported by the association of local and low-$z$ 
LGRBs with the explosion of a broad-line type Ic SN \citep[e.g.][]{dellaValle2003,hjo03,pia06,Bufano2012,Melandri2012}.
While the precise physical details of the LGRB phenomena are still
matter of debate, it is reasonable to assume that the probability for a
LGRB to be hosted in a given galaxy should be proportional to the mass
of young stellar particles (with ages below $\sim
10^7$ yr, corresponding to stellar masses $\simgt 30$ M$_\odot$) in each galaxy.
Here we neglect the LGRBs powered by Pop III stars, whose
contribution is likely to be sub-dominant with respect to the Pop II
population  in the redshift range considered
\citep{campisi2011}. This choice is further motivated by the fact that
observed high-$z$ LGRBs are qualitatively similar to low- and
intermediate-$z$ ones suggesting common progenitors \citep{sal09}.
Additionally, in selecting galaxy hosts we do not consider any metallicity bias for the LGRB formation.
Indeed, \cite{campisi2011b} have shown that strong cut-off in the progenitor metallicity of the order of $ 0.1-0.3~\Zsun $ are at odd with the consistency between LGRB host galaxies at $z<1$ and the observed 
Fundamental Metallicity Relation (FMR; \citealt{Mannucci2011}).  While
  higher critical metallicities for the LGRB progenitors \cite[as
  suggested for example by][]{Nuza2007,Georgy2009} can not be excluded
  at the present stage, we want to stress that our results 
  are not affected by the existence of such critical
  threshold. Indeed, the large majority of the young stellar particles
  from which we compute the probability of the galaxy to host a LGRB event
  have stellar metallicities $Z<0.3\;\Zsun$ in the redshift range
  considered. Therefore, our results do not represent in any
  way a test for the existence of a metallicity bias for the LGRB formation.

\subsection{LGRB host distribution and SFR}

A quantitative estimate of the probability of finding a LGRB in a given galaxy can be simply obtained studying the SFRs.
At a first glance, one would think that larger objects have larger SFRs
and hence a larger probability of forming LGRBs. However, while this holds for a single galaxy, this statement is not necessarily true on a
global scale, as it does not take into account the lower cosmic
abundance of bigger structures.
Thus, we measure the statistical probability of LGRB hosts by
considering a sample formed by the three simulations described in      
Section~ \ref{sect:simulations} and computing the differential
probability distribution of LGRBs, $\di P_{GRB}$, per logarithmic SFR:
\begin{equation}
\label{eq:dPdLogSFR}
\frac{\di P_{GRB}}{\rm \di log(SFR) } =
\frac{\di N_{GRB} ({\rm log(SFR))}} {N_{GRB} \, \rm \di log(SFR) },
\end{equation}
where, $ \di N_{GRB} ({\rm log(SFR)})$ is the number of LGRB expected in each SFR bin
(in units of M$_\odot$/yr), normalized by their total number, $N_{GRB}$.

We display the resulting distributions in Fig.~\ref{fig:mappedPGRB} for L30 (left panel), L10 (central)
and L5 (right).
In general, the SFRs vary over a wide range of values (see discussion in
Sec.~\ref{sub:properties}).
Typically, in L30 more massive objects with larger star formation rates are found, while in L5 small objects are resolved which are not present in the other two boxes. Intermediate values of the SFR are found in L10. 
From the Figure, it is clear that the most frequent values for the LGRB probability defined in eq.~(\ref{eq:dPdLogSFR}) are around or below $\sim 0.15$ (blue) in L30, compared to $\sim 0.15 - 0.30$ (between blue and yellow) in L10, and $ \simgt 0.30$ (yellow and light yellow) in L5.
This shows that, contrary to what naively expected, LGRBs are more probably found in intermediate- or small-size objects with lower SFRs, and not in (often highly star forming) large systems, due to their paucity in the cosmological landscape.
This is particularly relevant for high-redshift detections and
investigations of primordial stars and also
suggests that LGRB host luminosities should be quite dim (see next Section).

\subsection{Luminosities of high-$z$ LGRB host galaxies}

\begin{figure}
  \vspace{-1.2cm}
\center{\includegraphics[scale=0.44]{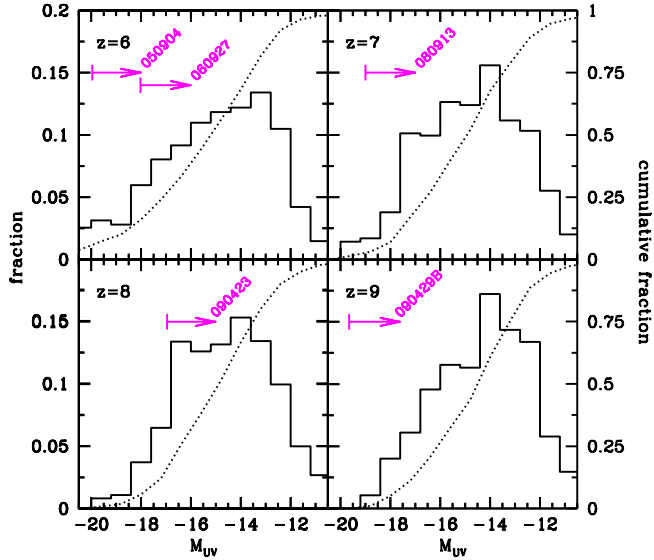}}
\caption{
Distributions of LGRB host absolute magnitude $M_{UV}$ at $z=6$, 7, 8, and 9, as indicated by the legends. Arrows show the lower limits obtained by Tanvir et al. (2012), with the corresponding LGRB name. In each panel, the solid line shows the differential distribution (scale on the left y-axis), while the dotted line shows the normalized cumulative distribution (scale on the right y-axis).
}
\label{fig:Muv}
\end{figure}

The distribution of UV luminosities of our simulated hosts is shown in
Fig.~\ref{fig:Muv}, together with the limits obtained in the deepest
HST/WFC3 observations by \cite{Tanvir2012}. As explained in previous
Sections, the sample is
obtained by jointly considering the results of our three simulation boxes.
At $z=6$ the probability shows a wide distribution, but only a very small fraction of LGRBs would reside in galaxies brighter than $M_{UV}=-18$, i.e. accessible to current instruments.
This reflects the fact that most of the young stars form in faint galaxies. At higher redshifts, the peak of the distribution shifts towards fainter galaxies, and mimics the evolution of the characteristic luminosity of the galaxy LF. In particular, at $z=8$ only $\sim 10$\% of LGRB hosts would be brighter than the extremely deep limit obtained for LGRB 090423. The non-detection of LGRB hosts at $z>6$ is then supporting the idea that most of the star formation arises in the faint-end of the galaxy LF at those early epochs, when mini-haloes host the initial phases of primordial star formation.
Our results also support the hypothesis that LGRB host galaxies are
likely to be too faint to be detected  by currently available space-
and ground-based facilities. Deep JWST observations are therefore
needed to image high-$z$ LGRB host galaxies. We find that in order to
have a 50\% probability to detect a LGRB host at $z=6$ at 5$\sigma$
level, an observation of $\sim 2\times 10^5$ sec should be foreseen
with the F150W filter on NIRCAM\footnote{On the base of current
  version of the JWST exposure time calculator,
  http://jwstetc.stsci.edu/etc/.}. For $z=8$ LGRBs, $\sim 3$ times larger
integration times are expected.

\begin{figure}
  \center{\includegraphics[scale=0.4]{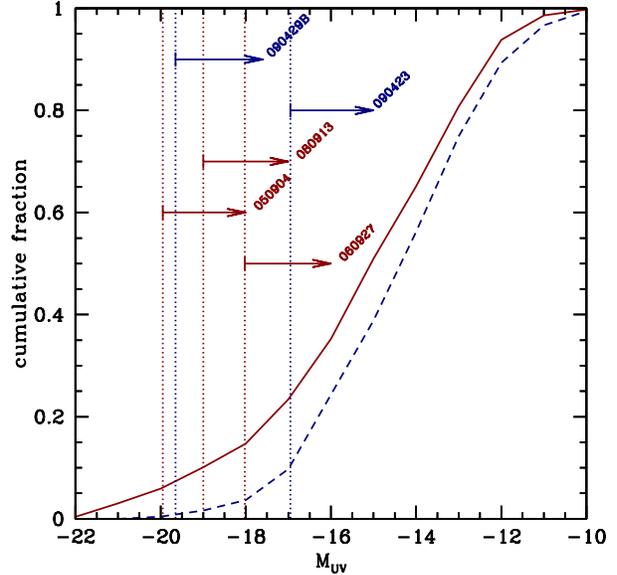}}
  \caption{
    Fraction of ionizing photons produced in galaxies fainter than $M_{UV}$. Arrows show the lower limits obtained by Tanvir et al. (2012) with the corresponding GRB name. Solid (dashed) line refers to $z=6$ ($z=8$) galaxies.}
  \label{fig:reion}
\end{figure}

These results suggest that LGRBs can be used as signpost for those very faint, but extremely common, galaxies that are now thought to reionize the Universe.
This is further illustrated in Fig.~\ref{fig:reion}, where we show the fraction of ionizing photons produced in galaxies fainter than a given absolute magnitude M$_{UV}$.
Most of the ionizing photons are provided by galaxies much fainter than those resolved by HST/WFC3 observations and will be resolved only in the deepest JWST fields. The limits on the luminosities of high-$z$ LGRB hosts are over-plotted in the Figure. 
Less than 10\% of the ionizing photons are provided by galaxies brighter than the limits reached by current LGRB host searches.
This holds true even in the case of GRB~090423, in spite of the extreme limit reached by the dedicated HST observations.
Thus, high-$z$ LGRBs provide the first (best) targets for JWST and ELT
observations aimed to seek the sources of cosmic reionization
\cite[see also][]{Trenti2012}.

\subsection{Statistical properties of high-$z$ LGRB host galaxies}\label{sub:properties}

\begin{figure}
\centering
 {\includegraphics[scale=0.4]{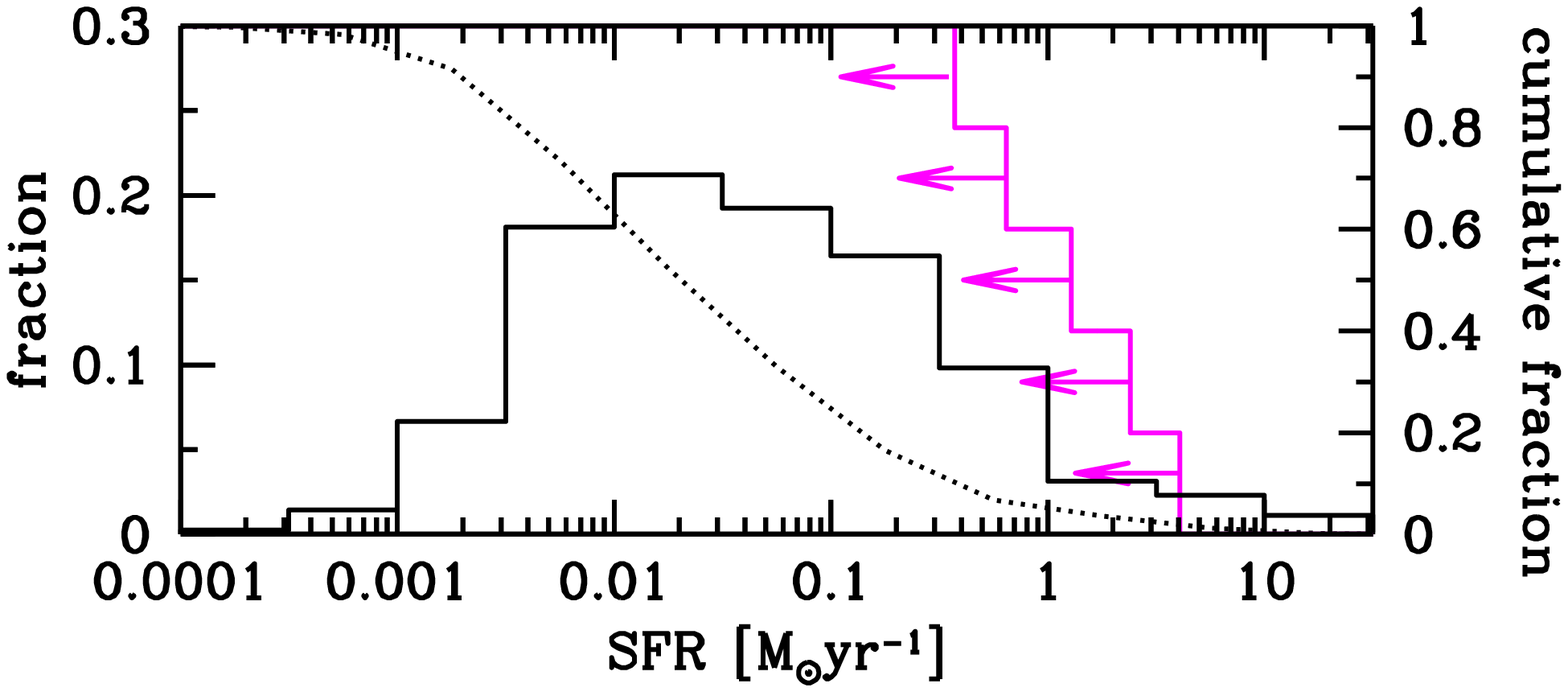}}\qquad
 {\includegraphics[scale=0.4]{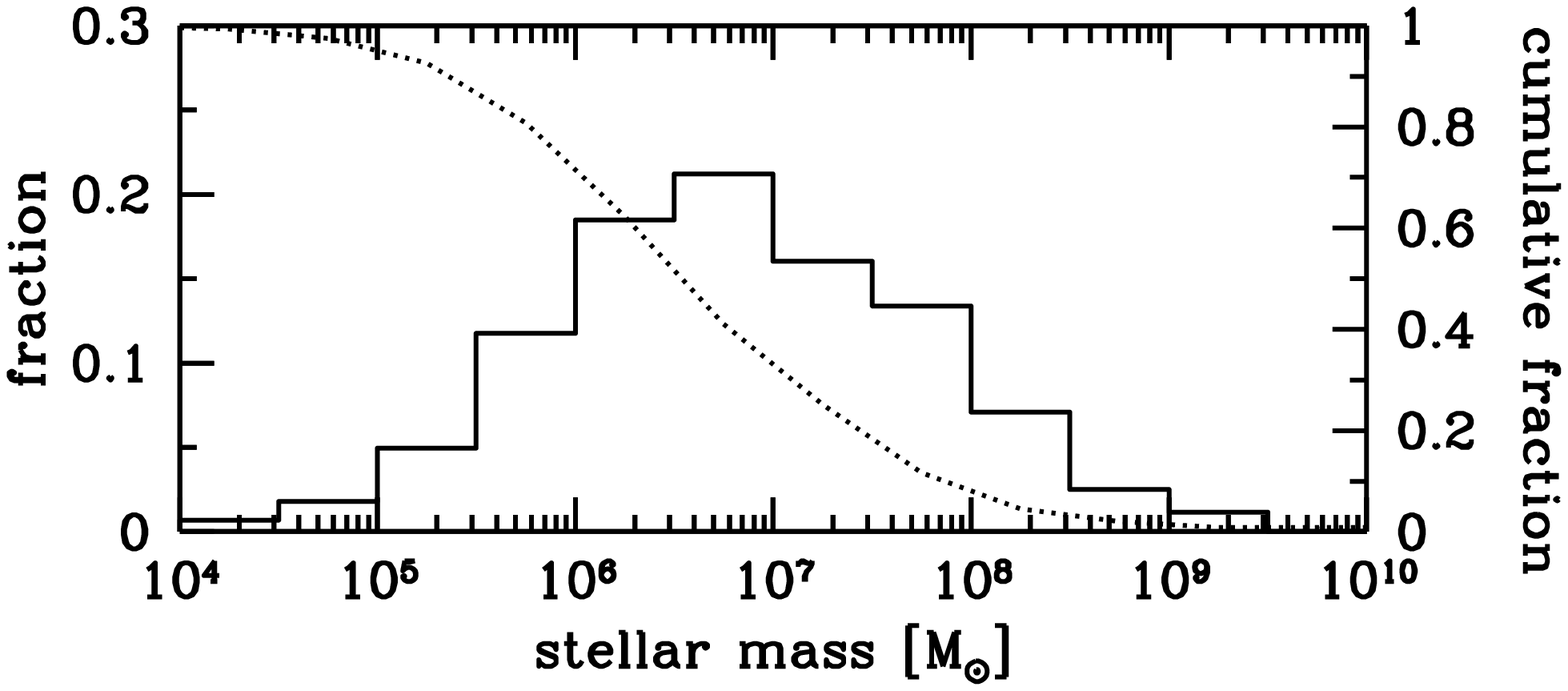}}
 {\includegraphics[scale=0.4]{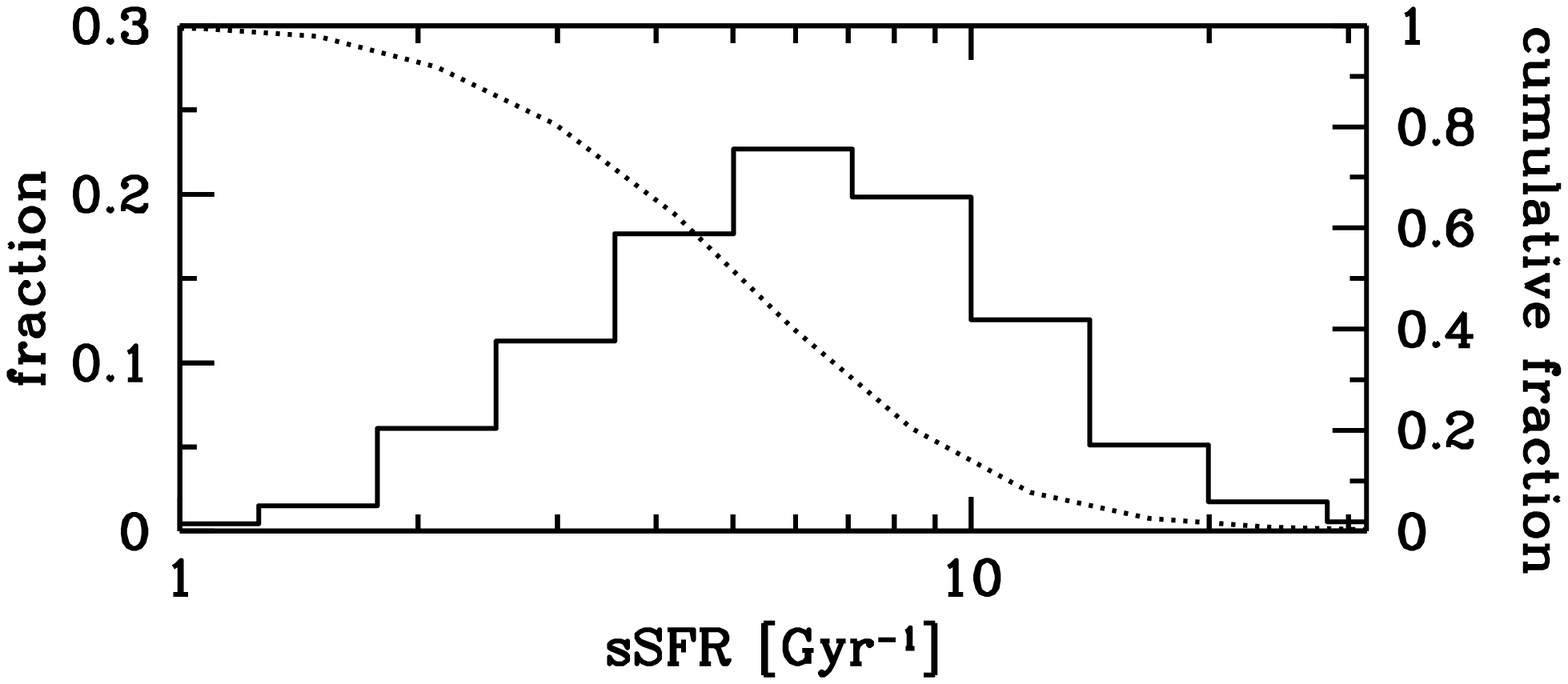}}
 {\includegraphics[scale=0.4]{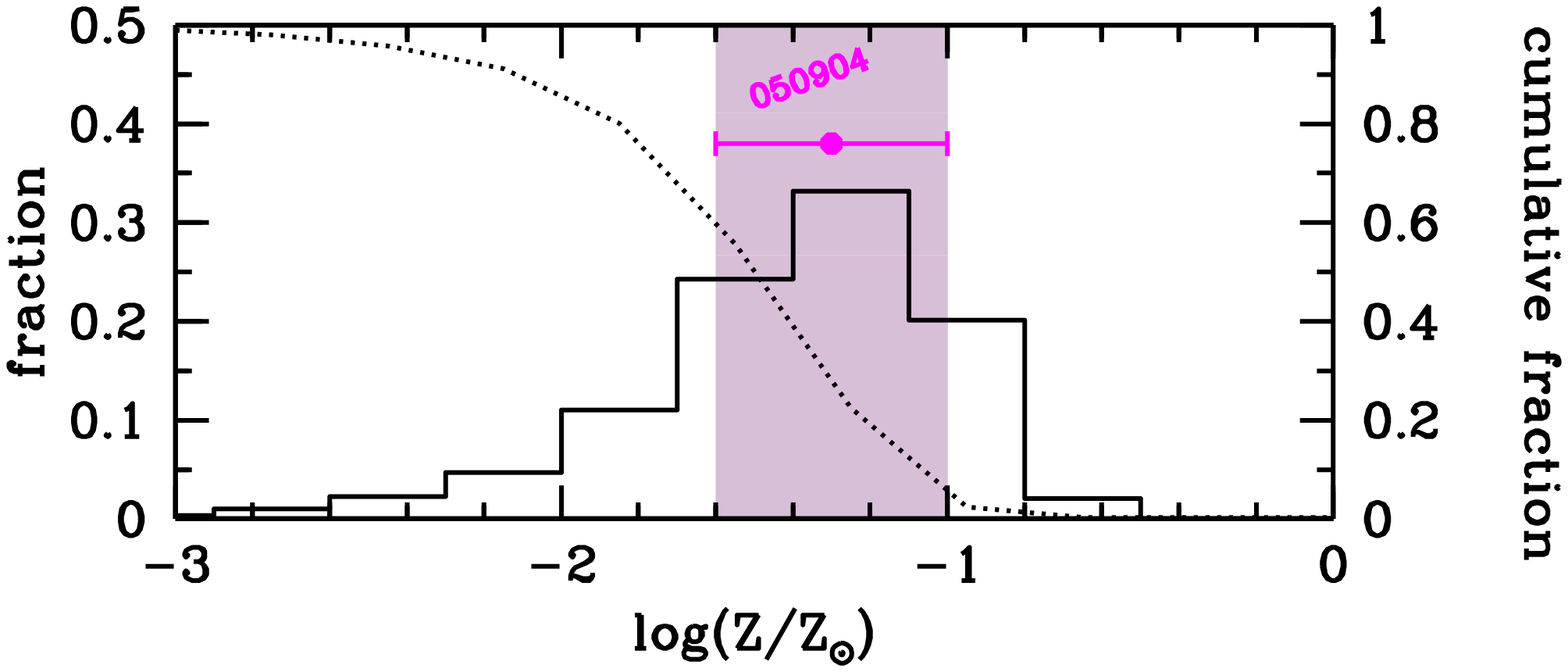}}
\caption{
Properties of simulated LGRB hosts in the redshift range
$z=6-10$. Panels from top to bottom show the distribution of SFR (in
$\rm \Msun~yr^{-1}$), stellar masses (in $\rm \Msun$), specific SFR
(in $\rm Gyr^{-1}$), and gas phase metallicities (in solar units, $Z/\Zsun$). The dotted line shows the normalized cumulative distribution (see right y-axis). The histogram with arrows in the top panel reports the limits on the SFR of LGRB hosts as obtained by Tanvir et al. (2012).
The data point in the bottom panel refers to the metallicity inferred from Si absorption lines in the afterglow spectrum of GRB~050904 at $z=6.3$, by Kawai et al. (2006).
}
\label{fig:prop}
\end{figure}

Now, we further explore the other physical properties of LGRB host galaxies in the redshift range $z=6-10$, for our whole simulated sample.
In Fig.~\ref{fig:prop}, from top to bottom, the panels show the
distribution of SFRs, stellar masses, specific SFR (sSFR, i.e. the SFR
per units of stellar mass) and  gas phase metallicities. In all panels the  dotted lines refer to the cumulative fraction.

The distribution of SFRs indicates that most of the simulated hosts have SFR $\sim 0.03-0.3~\Msun$ 
yr$^{-1}$ with only a few percent having $\rm SFR >1~\Msun~yr^{-1}$. This again is in line with the 
limits derived from individual observations of the fields of the most distant LGRBs by \cite{Tanvir2012}
and \cite{Basa2012}.
Indeed, considering current observational limits, the 
probability of non detection for all the five targets is as high as
70\% when the SFR distribution of simulated galaxies is
adopted. It is interesting to compare this value to the probability
of non detection expected assuming that $z>6$ galaxies have a 
distribution similar to that of LGRB hosts detected at $z<1$. Basa et al. (2012) report
that in this case the probability is only $\sim
6$\%, suggesting that the SFR distribution of high-$z$ hosts 
should peak at much lower values than
low-$z$ one, as indeed found in our simulations.
Assuming that LGRBs are good tracers of the formation of stars
\cite[but see e.g.][]{sal07,sal12}, current observational limits may
already provide the evidence that at high $z$ the bulk of stars resides in objects with much lower SFRs than in the local Universe, as also expected from numerical simulations of early structure formation \cite[e.g.][]{abel2002, yoshida2003, maio2010, wise2012}.

The majority of the simulated LGRB hosts is found to have relatively low stellar masses, $M_\star \sim 10^6-10^8~\Msun$.
As a result, in spite of their relatively low SFR, high-$z$ LGRB hosts have a very large sSFR, of $\rm \sim 3-10\,Gyr^{-1}$, showing that they are experiencing a strong burst of star formation.
The mean sSFR of high-$z$ hosts is much larger than the observed low-$z$ LGRB one \cite[estimated to be $\rm\sim 0.8~Gyr^{-1}$;][]{Savaglio2009}.
However, their doubling time (equal to the inverse of the sSFR), $t_{db}$, is $\sim 0.1-0.3$ of the Hubble time, $t_H(z)$, similar to what found for the low-$z$ LGRB hosts, i.e.
\begin{equation}
\rm  \frac{t_{db}(z)}{t_H(z)} = \frac{1}{sSFR(z)~t_H(z)} \sim 0.1-0.3
\end{equation}
at all redshifts.

In spite of the high $z$, LGRB hosts have gas phase metallicities in the range $-0.5<\log(Z/\Zsun)<-3$, with a peak in the distribution around $Z\sim 0.03\;\Zsun$.
These values are in line with those inferred from absorption line
measurements\footnote{Metal absorption lines
  over-imposed on the LGRB afterglow probe properly the metal content
  along the line-of-sight and not necessarily of the galaxy itself} in the LGRB afterglow at lower redshift, $2<z<4$, suggesting that the properties of the environment in which LGRBs explode are not so different in the early Universe.
This scenario is supported by the metallicity inferred from the analysis of the afterglow 
spectrum of GRB~050904 at $z=6.3$
(as marked in the bottom panel of Figure~\ref{fig:prop})\footnote{A slightly 
lower metallicity for GRB 050904 has been computed recently by \cite{Thoene2012}, 
log$(Z/\Zsun)=-1.6\pm 0.1$. This value is still consistent with our findings.}.
Besides that, no other metallicity constraint at $z>6$ has been
obtained so far owing to the low signal-to-noise ratio of the
spectra. 
However, we note that other LGRBs detected at $z=4-5$, e.g. GRB~090205 \citep{DAvanzo2010} 
and GRB~100219A \citep{Thoene2012}, show metallicities $Z\ge 0.1\;\Zsun$, further supporting our findings.
This is also in line with numerical studies of cosmic metal enrichment, that predict efficient pollution from early stellar populations \citep{tornatore2007, maio2010, maio2011,wise2012}, and the consequent presence of significative amounts of metals in high-$z$ LGRB hosts \citep{campisi2011}.

\begin{figure*}
\vspace{-0.7cm}
\centering
\includegraphics[width=0.7\textwidth, angle=-90]{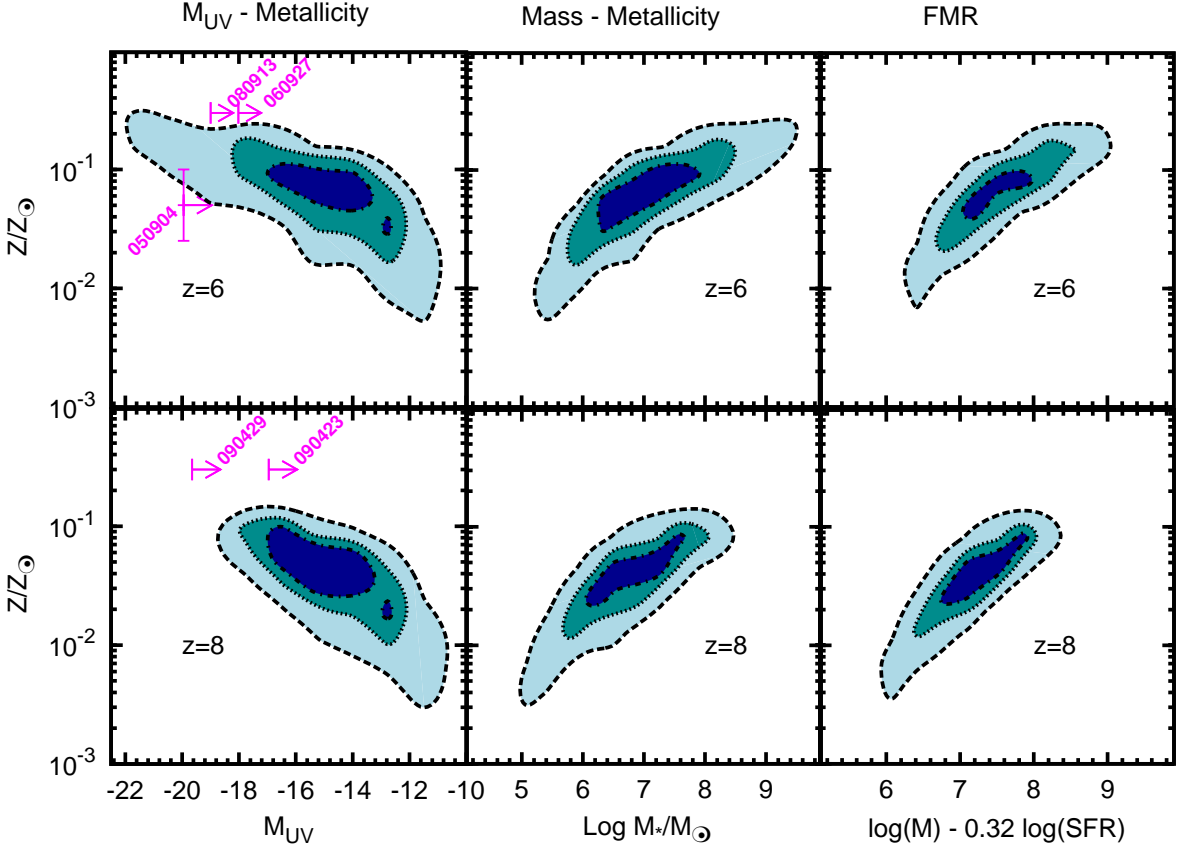}
\caption{
Luminosity-metallicity (left columns), mass-metallicity (central columns) and Fundamental Metallicity Relation (right columns)
for the LGRB host galaxies at $z=6$ (upper rows) and $z=8$ (lower rows).
Contour plots report the 30\%, 60\%, and 90\% probability of hosting a
LGRB. Arrows refer to Tanvir et al. (2012) results and, in the absence of a measured metallicity,
have been positioned arbitrarily at $Z=0.3\;\Zsun$, while the
metallicity of LGRB 050904 has been obtained by Kawai et al. (2006).
}
\label{fig:MZ} 
\end{figure*}

\subsection{Physical relations for LGRB host galaxies}

To further characterize the environment properties of high-$z$ LGRBs,
we compute the luminosity-metallicity relation (M$_{UV}$-Z), mass-metallicity relation (M-Z) 
and FMR relation of the LGRB hosts. 
These are shown in Fig.~\ref{fig:MZ}, for $z=6$ and $z=8$, where
contour levels refer to the loci in which we expect to find 30\%,
60\%, and 90\% of the simulated LGRBs.
We find that LGRB hosts, and simulated high-$z$ galaxies in general, follow well
defined relations in these planes. The luminosity-metallicity relation
shows the largest scatter that reduces significantly considering the
M-Z relation. An even tighter relation is found when the parameter
$\mu_{0.32}=\log(M_\star)-0.32\log({\rm SFR})$ is considered. 
This trend is similar to what
found for field galaxies  at lower redshift \citep{Mannucci2010}
and for LGRB hosts at $z<1$ \citep{Mannucci2011}. 
As expected, LGRBs are tracing the low-mass faint-end of the relations.
Measurements of metallicities in high-$z$ LGRB afterglows can therefore provide
unique information about the metal enrichment history of very small
galaxies in the primordial Universe. 

Adopting the M$_{UV}$-Z relation
and considering the observational limits on the LGRB host luminosities, we can
infer that the bursts detected so far at $z>6$  should have exploded in
galaxies with $Z<0.1\;\Zsun$. Moreover, we find that the host galaxy
of GRB~050904, in order to be consistent with the simulated relation, should be at
least two magnitudes fainter than current upper limits. Thus, its
detection would result very difficult even in extremely long
HST observations as done in the case of GRB~090423.

\cite{Thoene2012} presented a sample of LGRBs at $z=4-5$ for which the measure of the 
metallicity and of the host galaxy luminosity has been obtained. Four out of the six LGRBs 
considered by \cite{Thoene2012} show metallicities $\gsim 0.1\;\Zsun$. The corresponding 
LGRB host luminosities in the rest-frame B band are in the range [-22,-20]. The other two 
LGRBs have $Z<0.1\;\Zsun$, but only upper limits on the host luminosity are given, with M$_B\gsim -21$. 
Although these bursts lie at redshifts slightly lower than those studied here, we note that the 
position of these bursts in the L-Z plane is consistent with our findings. In particular, 
they support the idea that LGRB host galaxies should be already enriched at the level of a 
few percent of the solar metallicity at $z\ge 6$.

It is interesting to compare these maps with the analogous for Pop~III LGRB hosts, as computed by \cite{campisi2011}. 
With respect to Pop~III LGRB hosts, we find that Pop~II LGRB hosts are more massive and more metal enriched at all redshifts.
Indeed, most of the Pop~III LGRBs explode in galaxies near the critical metallicity line.
Although some is found to be hosted in the outer edge of more evolved galaxies -- see Figure~3 in \cite{campisi2011}.
Low metal abundances and peculiar abundance ratios may help in discriminating between Pop~III and Pop~II LGRBs.

\section{Conclusions}\label{sect:conclusions}

We investigated the nature of high-$z$ host galaxies of LGRBs by
means of N-body, hydrodynamical, chemistry simulations
\citep{maio2010} of galaxy formation and evolution, tracing both
Population III and Population II-I star forming regimes, and providing
a good description of the general field galaxy population at $z>6$
in the luminosity range probed by current HST observations. 

We predicted the physical properties of the LGRB host population in the
light of our understanding of galaxy formation and evolution in the
early Universe. 
We found that high-redshift ($z\sim 6-10$) LGRBs are hosted in bursty galaxies with typical star formation rates $\rm SFR\simeq 0.03-0.3\;\Msun~ yr^{-1}$, stellar masses $M_\star\simeq 10^6-10^8~\Msun$, specific star formation rates $\rm sSFR \simeq 3-10\,Gyr^{-1}$ (larger than the low-$z$ counterparts of up to one order of magnitude), and metallicities $Z\simeq 0.01-0.1\;\Zsun$.
Furthermore, the ratio between the resulting LGRB host doubling time
and the corresponding cosmic time seems to be universally equal to
roughly $\sim 0.1-0.3$, independently from the redshift.

The distribution of their UV luminosity places simulated LGRB hosts in the
faint-end of the galaxy luminosity function, well below the current
capabilities of space- or ground-based optical facilities, in line
with recent reports of non-detection of LGRB hosts in extremely deep
HST and VLT observations. Deep JWST observations should be foreseen in
order to detect these objects. These results suggest that LGRBs are the
signposts of those faint, but extremely common, galaxies that are
thought to reionize the Universe.


\section*{Acknowledgements}
RS and BC gratefully acknowledge all members of the DAVID collaboration (www.arcetri.astro.it/david) for inspiration and stimulating discussion. RS thanks S.~Basa for useful discussions about high-$z$ LGRB host observations.
UM thanks S.~Berta, D.~Pierini, and P.~Schady for useful discussions. He also acknowledges the Garching Computing Center of the Max Planck Society (Rechenzentrum Garching, RZG) for invaluable technical support, and financial contributions from the Project HPC-Europa2, grant number 228398, with the support of the European Community, under the FP7 Research Infrastructure Programme.
In this paper, we made use of the tools offered by the NASA Astrophysics Data System for the bibliographic research.

\bibliographystyle{mn2e}
\bibliography{host.bib}

\begin{thebibliography}{}

\bibitem[\protect\citeauthoryear{{Abel}, {Bryan} \& {Norman}}{{Abel}
  et~al.}{2002}]{abel2002}
{Abel} T.,  {Bryan} G.~L.,    {Norman} M.~L.,  2002, Science, 295, 93

\bibitem[\protect\citeauthoryear{{Baraffe}, {Heger} \& {Woosley}}{{Baraffe}
  et~al.}{2001}]{baraffe2001}
{Baraffe} I.,  {Heger} A.,    {Woosley} S.~E.,  2001, \apj, 550, 890

\bibitem[\protect\citeauthoryear{{Basa}, {Cuby}, {Savaglio}, {Boissier},
  {Cl{\'e}ment}, {Flores}, {Le Borgne} \& {Mazure}}{{Basa}
  et~al.}{2012}]{Basa2012}
{Basa} S.,  {Cuby} J.~G.,  {Savaglio} S.,  {Boissier} S.,  {Cl{\'e}ment} B.,
  {Flores} H.,  {Le Borgne} D.,    {Mazure} A.,  2012, \aap, 542, A103

\bibitem[\protect\citeauthoryear{{Bate} \& {Burkert}}{{Bate} \&
  {Burkert}}{1997}]{Bate1997}
{Bate} M.~R.,  {Burkert} A.,  1997, \mnras, 288, 1060

\bibitem[\protect\citeauthoryear{{Bouwens}, {Illingworth}, {Franx} \&
  {Ford}}{{Bouwens} et~al.}{2007}]{Bouwens2007}
{Bouwens} R.~J.,  {Illingworth} G.~D.,  {Franx} M.,    {Ford} H.,  2007, \apj,
  670, 928

\bibitem[\protect\citeauthoryear{{Bouwens}, {Illingworth}, {Oesch},
  {Labb{\'e}}, {Trenti}, {van Dokkum}, {Franx}, {Stiavelli}, {Carollo}, {Magee}
  \& {Gonzalez}}{{Bouwens} et~al.}{2011}]{Bouwens2011}
{Bouwens} R.~J.,  {Illingworth} G.~D.,  {Oesch} P.~A.,  {Labb{\'e}} I.,
  {Trenti} M.,  {van Dokkum} P.,  {Franx} M.,  {Stiavelli} M.,  {Carollo}
  C.~M.,  {Magee} D.,    {Gonzalez} V.,  2011, \apj, 737, 90

\bibitem[\protect\citeauthoryear{{Bromm} \& {Loeb}}{{Bromm} \&
  {Loeb}}{2003}]{Bromm2003}
{Bromm} V.,  {Loeb} A.,  2003, \nat, 425, 812

\bibitem[\protect\citeauthoryear{{Bufano}, {Pian}, {Sollerman}, {Benetti},
  {Pignata}, {Valenti}, {Covino} \& {et al.}}{{Bufano}
  et~al.}{2012}]{Bufano2012}
{Bufano} F.,  {Pian} E.,  {Sollerman} J.,  {Benetti} S.,  {Pignata} G.,
  {Valenti} S.,  {Covino} S.,    {et al.} 2012, \apj, 753, 67

\bibitem[\protect\citeauthoryear{{Campisi}, {De Lucia}, {Li}, {Mao} \&
  {Kang}}{{Campisi} et~al.}{2009}]{Campisi2009}
{Campisi} M.~A.,  {De Lucia} G.,  {Li} L.-X.,  {Mao} S.,    {Kang} X.,  2009,
  \mnras, 400, 1613

\bibitem[\protect\citeauthoryear{{Campisi}, {Maio}, {Salvaterra} \&
  {Ciardi}}{{Campisi} et~al.}{2011a}]{campisi2011}
{Campisi} M.~A.,  {Maio} U.,  {Salvaterra} R.,    {Ciardi} B.,  2011a, \mnras,
  416, 2760

\bibitem[\protect\citeauthoryear{{Campisi}, {Tapparello}, {Salvaterra},
  {Mannucci} \& {Colpi}}{{Campisi} et~al.}{2011b}]{campisi2011b}
{Campisi} M.~A.,  {Tapparello} C.,  {Salvaterra} R.,  {Mannucci} F.,    {Colpi}
  M.,  2011b, \mnras, 417, 1013

\bibitem[\protect\citeauthoryear{{Chisari}, {Tissera} \& {Pellizza}}{{Chisari}
  et~al.}{2010}]{Chisari2010}
{Chisari} N.~E.,  {Tissera} P.~B.,    {Pellizza} L.~J.,  2010, \mnras, 408, 647

\bibitem[\protect\citeauthoryear{{Ciardi} \& {Loeb}}{{Ciardi} \&
  {Loeb}}{2000}]{ciardi00}
{Ciardi} B.,  {Loeb} A.,  2000, \apj, 540, 687

\bibitem[\protect\citeauthoryear{{Courty}, {Bj{\"o}rnsson} \&
  {Gudmundsson}}{{Courty} et~al.}{2004}]{Courty2004}
{Courty} S.,  {Bj{\"o}rnsson} G.,    {Gudmundsson} E.~H.,  2004, \mnras, 354,
  581

\bibitem[\protect\citeauthoryear{{Cucchiara}, {Levan}, {Fox}, {Tanvir},
  {Ukwatta}, {Berger}, {Kr{\"u}hler}, {K{\"u}pc{\"u} Yolda{\c s}} \& {et
  al.}}{{Cucchiara} et~al.}{2011}]{Cucchiara2011}
{Cucchiara} A.,  {Levan} A.~J.,  {Fox} D.~B.,  {Tanvir} N.~R.,  {Ukwatta}
  T.~N.,  {Berger} E.,  {Kr{\"u}hler} T.,  {K{\"u}pc{\"u} Yolda{\c s}} A.,
  {et al.} 2011, \apj, 736, 7

\bibitem[\protect\citeauthoryear{{D'Avanzo}, {Perri}, {Fugazza}, {Salvaterra},
  {Chincarini}, {Margutti}, {Wu} \& {et al.}}{{D'Avanzo}
  et~al.}{2010}]{DAvanzo2010}
{D'Avanzo} P.,  {Perri} M.,  {Fugazza} D.,  {Salvaterra} R.,  {Chincarini} G.,
  {Margutti} R.,  {Wu} X.~F.,    {et al.} 2010, \aap, 522, A20

\bibitem[\protect\citeauthoryear{{Dayal}, {Ferrara} \& {Saro}}{{Dayal}
  et~al.}{2010}]{Dayal2010}
{Dayal} P.,  {Ferrara} A.,    {Saro} A.,  2010, \mnras, 402, 1449

\bibitem[\protect\citeauthoryear{{de Souza}, {Yoshida} \& {Ioka}}{{de Souza}
  et~al.}{2011}]{deSouza2011}
{de Souza} R.~S.,  {Yoshida} N.,    {Ioka} K.,  2011, \aap, 533, A32

\bibitem[\protect\citeauthoryear{{Della Valle}, {Malesani}, {Benetti}, {Testa},
  {Hamuy}, {Antonelli}, {Chincarini}, {Cocozza}, {Covino}, {D'Avanzo},
  {Fugazza}, {Ghisellini}, {Gilmozzi}, {Lazzati}, {Mason}, {Mazzali} \&
  {Stella}}{{Della Valle} et~al.}{2003}]{dellaValle2003}
{Della Valle} M.,  {Malesani} D.,  {Benetti} S.,  {Testa} V.,  {Hamuy} M.,
  {Antonelli} L.~A.,  {Chincarini} G.,  {Cocozza} G.,  {Covino} S.,  {D'Avanzo}
  P.,  {Fugazza} D.,  {Ghisellini} G.,  {Gilmozzi} R.,  {Lazzati} D.,  {Mason}
  E.,  {Mazzali} P.,    {Stella} L.,  2003, \aap, 406, L33

\bibitem[\protect\citeauthoryear{{Dolag}, {Borgani}, {Murante} \&
  {Springel}}{{Dolag} et~al.}{2009}]{Dolag2009}
{Dolag} K.,  {Borgani} S.,  {Murante} G.,    {Springel} V.,  2009, \mnras, 399,
  497

\bibitem[\protect\citeauthoryear{{Dolag}, {Jubelgas}, {Springel}, {Borgani} \&
  {Rasia}}{{Dolag} et~al.}{2004}]{dolag2004}
{Dolag} K.,  {Jubelgas} M.,  {Springel} V.,  {Borgani} S.,    {Rasia} E.,
  2004, \apjl, 606, L97

\bibitem[\protect\citeauthoryear{{Fryer}, {Woosley} \& {Hartmann}}{{Fryer}
  et~al.}{1999}]{Fryer_Woosley_Hartmann_1999}
{Fryer} C.~L.,  {Woosley} S.~E.,    {Hartmann} D.~H.,  1999, \apj, 526, 152

\bibitem[\protect\citeauthoryear{{Gallerani}, {Salvaterra}, {Ferrara} \&
  {Choudhury}}{{Gallerani} et~al.}{2008}]{Gallerani2008}
{Gallerani} S.,  {Salvaterra} R.,  {Ferrara} A.,    {Choudhury} T.~R.,  2008,
  \mnras, 388, L84

\bibitem[\protect\citeauthoryear{{Galli} \& {Palla}}{{Galli} \&
  {Palla}}{1998}]{GalliPalla1998}
{Galli} D.,  {Palla} F.,  1998, \aap, 335, 403

\bibitem[\protect\citeauthoryear{{Georgy}, {Meynet}, {Walder}, {Folini} \&
  {Maeder}}{{Georgy} et~al.}{2009}]{Georgy2009}
{Georgy} C.,  {Meynet} G.,  {Walder} R.,  {Folini} D.,    {Maeder} A.,  2009,
  \aap, 502, 611

\bibitem[\protect\citeauthoryear{{Gnedin} \& {Abel}}{{Gnedin} \&
  {Abel}}{2001}]{gnedinabel2001}
{Gnedin} N.~Y.,  {Abel} T.,  2001, \na, 6, 437

\bibitem[\protect\citeauthoryear{{Greiner}, {Kr{\"u}hler}, {Fynbo}, {Rossi},
  {Schwarz}, {Klose}, {Savaglio}, {Tanvir}, {McBreen} \& {et al.}}{{Greiner}
  et~al.}{2009}]{Greiner09}
{Greiner} J.,  {Kr{\"u}hler} T.,  {Fynbo} J.~P.~U.,  {Rossi} A.,  {Schwarz} R.,
   {Klose} S.,  {Savaglio} S.,  {Tanvir} N.~R.,  {McBreen} S.,    {et al.}
  2009, \apj, 693, 1610

\bibitem[\protect\citeauthoryear{{Heger}, {Fryer}, {Woosley}, {Langer} \&
  {Hartmann}}{{Heger} et~al.}{2003}]{heger2003}
{Heger} A.,  {Fryer} C.~L.,  {Woosley} S.~E.,  {Langer} N.,    {Hartmann}
  D.~H.,  2003, \apj, 591, 288

\bibitem[\protect\citeauthoryear{{Heger} \& {Woosley}}{{Heger} \&
  {Woosley}}{2002}]{heger2002}
{Heger} A.,  {Woosley} S.~E.,  2002, \apj, 567, 532

\bibitem[\protect\citeauthoryear{{Heger} \& {Woosley}}{{Heger} \&
  {Woosley}}{2010}]{heger2010}
{Heger} A.,  {Woosley} S.~E.,  2010, \apj, 724, 341

\bibitem[\protect\citeauthoryear{{Hernquist} \& {Katz}}{{Hernquist} \&
  {Katz}}{1989}]{hernquist1989}
{Hernquist} L.,  {Katz} N.,  1989, \apjs, 70, 419

\bibitem[\protect\citeauthoryear{{Hernquist} \& {Springel}}{{Hernquist} \&
  {Springel}}{2003}]{hernquist2003}
{Hernquist} L.,  {Springel} V.,  2003, \mnras, 341, 1253

\bibitem[\protect\citeauthoryear{{Hjorth {et al.}}}{{Hjorth {et
  al.}}}{2003}]{hjo03}
{Hjorth {et al.}} 2003, Nature, 423, 847

\bibitem[\protect\citeauthoryear{{Inoue}, {Salvaterra}, {Choudhury}, {Ferrara},
  {Ciardi} \& {Schneider}}{{Inoue} et~al.}{2010}]{Inoue2010}
{Inoue} S.,  {Salvaterra} R.,  {Choudhury} T.~R.,  {Ferrara} A.,  {Ciardi} B.,
    {Schneider} R.,  2010, \mnras, 404, 1938

\bibitem[\protect\citeauthoryear{{Jaacks}, {Choi}, {Nagamine}, {Thompson} \&
  {Varghese}}{{Jaacks} et~al.}{2012}]{Jaacks2012}
{Jaacks} J.,  {Choi} J.-H.,  {Nagamine} K.,  {Thompson} R.,    {Varghese} S.,
  2012, \mnras, 420, 1606

\bibitem[\protect\citeauthoryear{{Katz}}{{Katz}}{1992}]{katz1992}
{Katz} N.,  1992, \apj, 391, 502

\bibitem[\protect\citeauthoryear{{Kawai}, {Kosugi}, {Aoki}, {Yamada}, {Totani},
  {Ohta}, {Iye}, {Hattori} \& {et al.}}{{Kawai} et~al.}{2006}]{kaw06}
{Kawai} N.,  {Kosugi} G.,  {Aoki} K.,  {Yamada} T.,  {Totani} T.,  {Ohta} K.,
  {Iye} M.,  {Hattori} T.,    {et al.} 2006, \nat, 440, 184

\bibitem[\protect\citeauthoryear{{Lacey}, {Baugh}, {Frenk} \& {Benson}}{{Lacey}
  et~al.}{2011}]{Lacey2011}
{Lacey} C.~G.,  {Baugh} C.~M.,  {Frenk} C.~S.,    {Benson} A.~J.,  2011,
  \mnras, 412, 1828

\bibitem[\protect\citeauthoryear{{Maio}}{{Maio}}{2011}]{maiocqg2011}
{Maio} U.,  2011, Classical and Quantum Gravity, 28, 225015

\bibitem[\protect\citeauthoryear{{Maio}, {Ciardi}, {Dolag}, {Tornatore} \&
  {Khochfar}}{{Maio} et~al.}{2010}]{maio2010}
{Maio} U.,  {Ciardi} B.,  {Dolag} K.,  {Tornatore} L.,    {Khochfar} S.,  2010,
  \mnras, 407, 1003

\bibitem[\protect\citeauthoryear{{Maio}, {Ciardi}, {Yoshida}, {Dolag} \&
  {Tornatore}}{{Maio} et~al.}{2009}]{maio2009}
{Maio} U.,  {Ciardi} B.,  {Yoshida} N.,  {Dolag} K.,    {Tornatore} L.,  2009,
  \aap, 503, 25

\bibitem[\protect\citeauthoryear{{Maio}, {Dolag}, {Ciardi} \&
  {Tornatore}}{{Maio} et~al.}{2007}]{maio2007}
{Maio} U.,  {Dolag} K.,  {Ciardi} B.,    {Tornatore} L.,  2007, \mnras, 379,
  963

\bibitem[\protect\citeauthoryear{{Maio}, {Dolag}, {Meneghetti}, {Moscardini},
  {Yoshida}, {Baccigalupi}, {Bartelmann} \& {Perrotta}}{{Maio}
  et~al.}{2006}]{maio2006}
{Maio} U.,  {Dolag} K.,  {Meneghetti} M.,  {Moscardini} L.,  {Yoshida} N.,
  {Baccigalupi} C.,  {Bartelmann} M.,    {Perrotta} F.,  2006, \mnras, 373, 869

\bibitem[\protect\citeauthoryear{{Maio} \& {Iannuzzi}}{{Maio} \&
  {Iannuzzi}}{2011}]{maioiannuzzi2011}
{Maio} U.,  {Iannuzzi} F.,  2011, \mnras, 415, 3021

\bibitem[\protect\citeauthoryear{{Maio} \& {Khochfar}}{{Maio} \&
  {Khochfar}}{2012}]{maiokhochfar2012}
{Maio} U.,  {Khochfar} S.,  2012, \mnras, 421, 1113

\bibitem[\protect\citeauthoryear{{Maio}, {Khochfar}, {Johnson} \&
  {Ciardi}}{{Maio} et~al.}{2011}]{maio2011}
{Maio} U.,  {Khochfar} S.,  {Johnson} J.~L.,    {Ciardi} B.,  2011, \mnras,
  414, 1145

\bibitem[\protect\citeauthoryear{{Maio}, {Koopmans} \& {Ciardi}}{{Maio}
  et~al.}{2011}]{mkc2011}
{Maio} U.,  {Koopmans} L.~V.~E.,    {Ciardi} B.,  2011, \mnras, 412, L40

\bibitem[\protect\citeauthoryear{{Maio}, {Salvaterra}, {Moscardini} \&
  {Ciardi}}{{Maio} et~al.}{2012}]{Maio2012}
{Maio} U.,  {Salvaterra} R.,  {Moscardini} L.,    {Ciardi} B.,  2012, \mnras,
  426, 2078

\bibitem[\protect\citeauthoryear{{Mannucci}, {Cresci}, {Maiolino}, {Marconi} \&
  {Gnerucci}}{{Mannucci} et~al.}{2010}]{Mannucci2010}
{Mannucci} F.,  {Cresci} G.,  {Maiolino} R.,  {Marconi} A.,    {Gnerucci} A.,
  2010, \mnras, 408, 2115

\bibitem[\protect\citeauthoryear{{Mannucci}, {Salvaterra} \&
  {Campisi}}{{Mannucci} et~al.}{2011}]{Mannucci2011}
{Mannucci} F.,  {Salvaterra} R.,    {Campisi} M.~A.,  2011, \mnras, 414, 1263

\bibitem[\protect\citeauthoryear{{McLure}, {Dunlop}, {Cirasuolo}, {Koekemoer},
  {Sabbi}, {Stark}, {Targett} \& {Ellis}}{{McLure} et~al.}{2010}]{McLure}
{McLure} R.~J.,  {Dunlop} J.~S.,  {Cirasuolo} M.,  {Koekemoer} A.~M.,  {Sabbi}
  E.,  {Stark} D.~P.,  {Targett} T.~A.,    {Ellis} R.~S.,  2010, \mnras, 403,
  960

\bibitem[\protect\citeauthoryear{{McQuinn}, {Bloom}, {Grindlay}, {Band},
  {Barthelmy}, {Berger}, {Corsi} \& {et al.}}{{McQuinn}
  et~al.}{2009}]{McQuinn2009}
{McQuinn} M.,  {Bloom} J.~S.,  {Grindlay} J.,  {Band} D.,  {Barthelmy} S.~D.,
  {Berger} E.,  {Corsi} A.,    {et al.} 2009, in astro2010: The Astronomy and
  Astrophysics Decadal Survey Vol.~2010 of Astronomy, {In Situ Probes of the
  First Galaxies and Reionization: Gamma-ray Bursts}.
p.~199

\bibitem[\protect\citeauthoryear{{McQuinn}, {Lidz}, {Zaldarriaga}, {Hernquist}
  \& {Dutta}}{{McQuinn} et~al.}{2008}]{McQuinn2008}
{McQuinn} M.,  {Lidz} A.,  {Zaldarriaga} M.,  {Hernquist} L.,    {Dutta} S.,
  2008, \mnras, 388, 1101

\bibitem[\protect\citeauthoryear{{Melandri}, {Pian}, {Ferrero}, {D'Elia},
  {Walker}, {Ghirlanda}, {Covino}, {Amati} \& {et al.}}{{Melandri}
  et~al.}{2012}]{Melandri2012}
{Melandri} A.,  {Pian} E.,  {Ferrero} P.,  {D'Elia} V.,  {Walker} E.~S.,
  {Ghirlanda} G.,  {Covino} S.,  {Amati} L.,    {et al.} 2012, \aap, 547, A82

\bibitem[\protect\citeauthoryear{{Mu{\~n}oz} \& {Loeb}}{{Mu{\~n}oz} \&
  {Loeb}}{2011}]{Munoz2011}
{Mu{\~n}oz} J.~A.,  {Loeb} A.,  2011, \apj, 729, 99

\bibitem[\protect\citeauthoryear{{Nagamine}, {Springel} \&
  {Hernquist}}{{Nagamine} et~al.}{2004}]{nag04}
{Nagamine} K.,  {Springel} V.,    {Hernquist} L.,  2004, \mnras, 348, 435

\bibitem[\protect\citeauthoryear{{Nagamine}, {Zhang} \& {Hernquist}}{{Nagamine}
  et~al.}{2008}]{Nagamine2008}
{Nagamine} K.,  {Zhang} B.,    {Hernquist} L.,  2008, \apjl, 686, L57

\bibitem[\protect\citeauthoryear{{Nuza}, {Tissera}, {Pellizza}, {Lambas},
  {Scannapieco} \& {de Rossi}}{{Nuza} et~al.}{2007}]{Nuza2007}
{Nuza} S.~E.,  {Tissera} P.~B.,  {Pellizza} L.~J.,  {Lambas} D.~G.,
  {Scannapieco} C.,    {de Rossi} M.~E.,  2007, \mnras, 375, 665

\bibitem[\protect\citeauthoryear{{Oesch}, {Bouwens}, {Illingworth}, {Gonzalez},
  {Trenti}, {van Dokkum}, {Franx}, {Labbe}, {Carollo} \& {Magee}}{{Oesch}
  et~al.}{2012}]{Oesch2012}
{Oesch} P.~A.,  {Bouwens} R.~J.,  {Illingworth} G.~D.,  {Gonzalez} V.,
  {Trenti} M.,  {van Dokkum} P.~G.,  {Franx} M.,  {Labbe} I.,  {Carollo} C.~M.,
     {Magee} D.,  2012, ArXiv e-prints, arXiv:1201.0755

\bibitem[\protect\citeauthoryear{{Padovani} \& {Matteucci}}{{Padovani} \&
  {Matteucci}}{1993}]{Padovani1993}
{Padovani} P.,  {Matteucci} F.,  1993, \apj, 416, 26

\bibitem[\protect\citeauthoryear{{Pian {et al.}}}{{Pian {et
  al.}}}{2006}]{pia06}
{Pian {et al.}} 2006, \nat, 442, 1011

\bibitem[\protect\citeauthoryear{{Price}}{{Price}}{2012}]{price2012b}
{Price} D.~J.,  2012, Journal of Computational Physics, 231, 759

\bibitem[\protect\citeauthoryear{{Ruiz-Velasco}, {Swan}, {Troja}, {Malesani},
  {Fynbo}, {Starling}, {Xu}, {Aharonian} \& {et al.}}{{Ruiz-Velasco}
  et~al.}{2007}]{Ruiz07}
{Ruiz-Velasco} A.~E.,  {Swan} H.,  {Troja} E.,  {Malesani} D.,  {Fynbo}
  J.~P.~U.,  {Starling} R.~L.~C.,  {Xu} D.,  {Aharonian} F.,    {et al.} 2007,
  \apj, 669, 1

\bibitem[\protect\citeauthoryear{{Salvaterra}, {Campana}, {Vergani}, {Covino},
  {D'Avanzo}, {Fugazza}, {Ghirlanda}, {Ghisellini}, {Melandri}, {Nava},
  {Sbarufatti}, {Flores}, {Piranomonte} \& {Tagliaferri}}{{Salvaterra}
  et~al.}{2012}]{sal12}
{Salvaterra} R.,  {Campana} S.,  {Vergani} S.~D.,  {Covino} S.,  {D'Avanzo} P.,
   {Fugazza} D.,  {Ghirlanda} G.,  {Ghisellini} G.,  {Melandri} A.,  {Nava} L.,
   {Sbarufatti} B.,  {Flores} H.,  {Piranomonte} S.,    {Tagliaferri} G.,
  2012, \apj, 749, 68

\bibitem[\protect\citeauthoryear{{Salvaterra} \& {Chincarini}}{{Salvaterra} \&
  {Chincarini}}{2007}]{sal07}
{Salvaterra} R.,  {Chincarini} G.,  2007, \apjl, 656, L49

\bibitem[\protect\citeauthoryear{{Salvaterra}, {Della Valle}, {Campana},
  {Chincarini}, {Covino}, {D'Avanzo}, {Fernandez-Soto}, {Guidorzi} \& {et
  al}}{{Salvaterra} et~al.}{2009}]{sal09}
{Salvaterra} R.,  {Della Valle} M.,  {Campana} S.,  {Chincarini} G.,  {Covino}
  S.,  {D'Avanzo} P.,  {Fernandez-Soto} A.,  {Guidorzi} C.,    {et al} 2009,
  \nat, 461, 1258

\bibitem[\protect\citeauthoryear{{Salvaterra}, {Ferrara} \&
  {Dayal}}{{Salvaterra} et~al.}{2011}]{Salvaterra2010}
{Salvaterra} R.,  {Ferrara} A.,    {Dayal} P.,  2011, \mnras, 414, 847

\bibitem[\protect\citeauthoryear{{Savaglio}, {Glazebrook} \& {Le
  Borgne}}{{Savaglio} et~al.}{2009}]{Savaglio2009}
{Savaglio} S.,  {Glazebrook} K.,    {Le Borgne} D.,  2009, \apj, 691, 182

\bibitem[\protect\citeauthoryear{{Savaglio}, {Rau}, {Greiner}, {Kr{\"u}hler},
  {McBreen}, {Hartmann}, {Updike}, {Filgas}, {Klose}, {Afonso}, {Clemens},
  {K{\"u}pc{\"u} Yolda{\c s}}, {Olivares E.}, {Sudilovsky} \&
  {Szokoly}}{{Savaglio} et~al.}{2012}]{Savaglio2012}
{Savaglio} S.,  {Rau} A.,  {Greiner} J.,  {Kr{\"u}hler} T.,  {McBreen} S.,
  {Hartmann} D.~H.,  {Updike} A.~C.,  {Filgas} R.,  {Klose} S.,  {Afonso} P.,
  {Clemens} C.,  {K{\"u}pc{\"u} Yolda{\c s}} A.,  {Olivares E.} F.,
  {Sudilovsky} V.,    {Szokoly} G.,  2012, \mnras, 420, 627

\bibitem[\protect\citeauthoryear{{Schaerer}}{{Schaerer}}{2002}]{Schaerer2002}
{Schaerer} D.,  2002, \aap, 382, 28

\bibitem[\protect\citeauthoryear{{Schneider}, {Ferrara}, {Natarajan} \&
  {Omukai}}{{Schneider} et~al.}{2002}]{Schneider2002}
{Schneider} R.,  {Ferrara} A.,  {Natarajan} P.,    {Omukai} K.,  2002, \apj,
  571, 30

\bibitem[\protect\citeauthoryear{{Schneider}, {Ferrara}, {Salvaterra}, {Omukai}
  \& {Bromm}}{{Schneider} et~al.}{2003}]{Schneider2003}
{Schneider} R.,  {Ferrara} A.,  {Salvaterra} R.,  {Omukai} K.,    {Bromm} V.,
  2003, \nat, 422, 869

\bibitem[\protect\citeauthoryear{{Schneider}, {Omukai}, {Limongi}, {Ferrara},
  {Salvaterra}, {Chieffi} \& {Bianchi}}{{Schneider}
  et~al.}{2012}]{Schneider2012}
{Schneider} R.,  {Omukai} K.,  {Limongi} M.,  {Ferrara} A.,  {Salvaterra} R.,
  {Chieffi} A.,    {Bianchi} S.,  2012, \mnras, 423, L60

\bibitem[\protect\citeauthoryear{{Springel} \& {Hernquist}}{{Springel} \&
  {Hernquist}}{2003}]{springel2003}
{Springel} V.,  {Hernquist} L.,  2003, \mnras, 339, 289

\bibitem[\protect\citeauthoryear{{Su}, {Stiavelli}, {Oesch}, {Trenti},
  {Bergeron}, {Bradley}, {Carollo}, {Dahlen}, {Ferguson}, {Giavalisco},
  {Koekemoer}, {Lilly}, {Lucas}, {Mobasher}, {Panagia} \& {Pavlovsky}}{{Su}
  et~al.}{2011}]{Su2011}
{Su} J.,  {Stiavelli} M.,  {Oesch} P.,  {Trenti} M.,  {Bergeron} E.,  {Bradley}
  L.,  {Carollo} M.,  {Dahlen} T.,  {Ferguson} H.~C.,  {Giavalisco} M.,
  {Koekemoer} A.,  {Lilly} S.,  {Lucas} R.~A.,  {Mobasher} B.,  {Panagia} N.,
   {Pavlovsky} C.,  2011, \apj, 738, 123

\bibitem[\protect\citeauthoryear{{Tanvir}, {Fox}, {Levan}, {Berger},
  {Wiersema}, {Fynbo}, {Cucchiara} \& {et al.}}{{Tanvir} et~al.}{2009}]{tan09}
{Tanvir} N.~R.,  {Fox} D.~B.,  {Levan} A.~J.,  {Berger} E.,  {Wiersema} K.,
  {Fynbo} J.~P.~U.,  {Cucchiara} A.,    {et al.} 2009, \nat, 461, 1254

\bibitem[\protect\citeauthoryear{{Tanvir}, {Levan}, {Fruchter}, {Fynbo},
  {Hjorth}, {Wiersema}, {Bremer}, {Rhoads} \& {et al.}}{{Tanvir}
  et~al.}{2012}]{Tanvir2012}
{Tanvir} N.~R.,  {Levan} A.~J.,  {Fruchter} A.~S.,  {Fynbo} J.~P.~U.,  {Hjorth}
  J.,  {Wiersema} K.,  {Bremer} M.~N.,  {Rhoads} J.,    {et al.} 2012, \apj,
  754, 46

\bibitem[\protect\citeauthoryear{{Thielemann}, {Argast}, {Brachwitz}, {Hix},
  {H{\"o}flich}, {Liebend{\"o}rfer}, {Martinez-Pinedo}, {Mezzacappa}, {Panov}
  \& {Rauscher}}{{Thielemann} et~al.}{2003}]{thielemann2003}
{Thielemann} F.-K.,  {Argast} D.,  {Brachwitz} F.,  {Hix} W.~R.,  {H{\"o}flich}
  P.,  {Liebend{\"o}rfer} M.,  {Martinez-Pinedo} G.,  {Mezzacappa} A.,  {Panov}
  I.,    {Rauscher} T.,  2003, Nuclear Physics A, 718, 139

\bibitem[\protect\citeauthoryear{{Thoene}, {Fynbo}, {Goldoni}, {de Ugarte
  Postigo}, {Campana}, {Vergani}, {Covino} \& {et al.}}{{Thoene}
  et~al.}{2012}]{Thoene2012}
{Thoene} C.~C.,  {Fynbo} J.~P.~U.,  {Goldoni} P.,  {de Ugarte Postigo} A.,
  {Campana} S.,  {Vergani} S.~D.,  {Covino} S.,    {et al.} 2012, ArXiv
  e-prints, arXiv:1206.2337

\bibitem[\protect\citeauthoryear{{Toma}, {Sakamoto} \&
  {M{\'e}sz{\'a}ros}}{{Toma} et~al.}{2011}]{toma10}
{Toma} K.,  {Sakamoto} T.,    {M{\'e}sz{\'a}ros} P.,  2011, \apj, 731, 127

\bibitem[\protect\citeauthoryear{{Tornatore}, {Borgani}, {Dolag} \&
  {Matteucci}}{{Tornatore} et~al.}{2007}]{tornatoreBDM2007}
{Tornatore} L.,  {Borgani} S.,  {Dolag} K.,    {Matteucci} F.,  2007, \mnras,
  382, 1050

\bibitem[\protect\citeauthoryear{{Tornatore}, {Borgani}, {Springel},
  {Matteucci}, {Menci} \& {Murante}}{{Tornatore} et~al.}{2003}]{tornatore2003}
{Tornatore} L.,  {Borgani} S.,  {Springel} V.,  {Matteucci} F.,  {Menci} N.,
  {Murante} G.,  2003, \mnras, 342, 1025

\bibitem[\protect\citeauthoryear{{Tornatore}, {Borgani}, {Viel} \&
  {Springel}}{{Tornatore} et~al.}{2010}]{tornatore2010}
{Tornatore} L.,  {Borgani} S.,  {Viel} M.,    {Springel} V.,  2010, \mnras,
  402, 1911

\bibitem[\protect\citeauthoryear{{Tornatore}, {Ferrara} \&
  {Schneider}}{{Tornatore} et~al.}{2007}]{tornatore2007}
{Tornatore} L.,  {Ferrara} A.,    {Schneider} R.,  2007, \mnras, 382, 945

\bibitem[\protect\citeauthoryear{{Trenti}, {Perna}, {Levesque}, {Shull} \&
  {Stocke}}{{Trenti} et~al.}{2012}]{Trenti2012}
{Trenti} M.,  {Perna} R.,  {Levesque} E.~M.,  {Shull} J.~M.,    {Stocke} J.~T.,
   2012, \apjl, 749, L38

\bibitem[\protect\citeauthoryear{{van den Hoek} \& {Groenewegen}}{{van den
  Hoek} \& {Groenewegen}}{1997}]{vandenhoek1997}
{van den Hoek} L.~B.,  {Groenewegen} M.~A.~T.,  1997, \aaps, 123, 305

\bibitem[\protect\citeauthoryear{{Wang}, {Bromm}, {Greif}, {Stacy}, {Dai},
  {Loeb} \& {Cheng}}{{Wang} et~al.}{2012}]{Wang2012}
{Wang} F.~Y.,  {Bromm} V.,  {Greif} T.~H.,  {Stacy} A.,  {Dai} Z.~G.,  {Loeb}
  A.,    {Cheng} K.~S.,  2012, ArXiv e-prints, arXiv:1207.3879

\bibitem[\protect\citeauthoryear{{Whalen} \& {Norman}}{{Whalen} \&
  {Norman}}{2008}]{whalen2008}
{Whalen} D.,  {Norman} M.~L.,  2008, \apj, 673, 664

\bibitem[\protect\citeauthoryear{{Wise} \& {Abel}}{{Wise} \&
  {Abel}}{2007}]{wise2007}
{Wise} J.~H.,  {Abel} T.,  2007, \apj, 665, 899

\bibitem[\protect\citeauthoryear{{Wise}, {Turk}, {Norman} \& {Abel}}{{Wise}
  et~al.}{2012}]{wise2012}
{Wise} J.~H.,  {Turk} M.~J.,  {Norman} M.~L.,    {Abel} T.,  2012, \apj, 745,
  50

\bibitem[\protect\citeauthoryear{{Woosley} \& {Heger}}{{Woosley} \&
  {Heger}}{2006}]{woo06b}
{Woosley} S.~E.,  {Heger} A.,  2006, \apj, 637, 914

\bibitem[\protect\citeauthoryear{{Woosley}, {Heger} \& {Weaver}}{{Woosley}
  et~al.}{2002}]{woosley2002}
{Woosley} S.~E.,  {Heger} A.,    {Weaver} T.~A.,  2002, Reviews of Modern
  Physics, 74, 1015

\bibitem[\protect\citeauthoryear{{Woosley} \& {Weaver}}{{Woosley} \&
  {Weaver}}{1995}]{woosley1995}
{Woosley} S.~E.,  {Weaver} T.~A.,  1995, \apjs, 101, 181

\bibitem[\protect\citeauthoryear{{Yoshida}, {Abel}, {Hernquist} \&
  {Sugiyama}}{{Yoshida} et~al.}{2003}]{yoshida2003}
{Yoshida} N.,  {Abel} T.,  {Hernquist} L.,    {Sugiyama} N.,  2003, \apj, 592,
  645

\end{thebibliography}

\label{lastpage}

\end{document}